\documentclass[twocolumn]{revtex4-1}
\usepackage{amsmath}
\usepackage{amsfonts}
\usepackage{amssymb}
\usepackage{graphicx}

\usepackage{mathtools}%% for loading postscript figures
\usepackage{color}

\usepackage{enumitem} 

\usepackage{bm}

\usepackage{here}

\usepackage{hyperref}

\begin{document}

\title{Function-Space Based Solution Scheme for the Size-Modified Poisson-Boltzmann Equation in Full-Potential DFT}

\author{Stefan Ringe}
\affiliation{Chair for Theoretical Chemistry and Catalysis Research Center, Technische Universit{\"a}t M{\"u}nchen, Lichtenbergstr. 4, D-85747 Garching, Germany}

\author{Harald Oberhofer}
\email{harald.oberhofer@tum.de}
\affiliation{Chair for Theoretical Chemistry and Catalysis Research Center, Technische Universit{\"a}t M{\"u}nchen, Lichtenbergstr. 4, D-85747 Garching, Germany}

\author{Christoph Hille}
\affiliation{Chair for Theoretical Chemistry and Catalysis Research Center, Technische Universit{\"a}t M{\"u}nchen, Lichtenbergstr. 4, D-85747 Garching, Germany}

\author{Sebastian Matera}
\affiliation{Fachbereich f. Mathematik u. Informatik, Freie Universit\"at Berlin, Otto-von-Simson-Str. 19, D-14195 Berlin, Germany}

\author{Karsten Reuter}
\affiliation{Chair for Theoretical Chemistry and Catalysis Research Center, Technische Universit{\"a}t M{\"u}nchen, Lichtenbergstr. 4, D-85747 Garching, Germany}

\date{\today}

\begin{abstract}
The size-modified Poisson-Boltzmann (MPB) equation is an efficient implicit solvation model which also captures electrolytic solvent effects. It combines an account of the dielectric solvent response with a mean-field description of solvated finite-sized ions. We present a general solution scheme for the MPB equation based on a fast function-space oriented Newton method and a Green's function preconditioned iterative linear solver. In contrast to popular multi-grid solvers this approach allows to fully exploit specialized integration grids and optimized integration schemes. We describe a corresponding numerically efficient implementation for the full-potential density-functional theory (DFT) code FHI-aims. We show that together with an additional Stern layer correction the DFT+MPB approach can describe the mean activity coefficient of a KCl aqueous solution over a wide range of concentrations. The high sensitivity of the calculated activity coefficient on the employed ionic parameters thereby suggests to use extensively tabulated experimental activity coefficients of salt solutions for a systematic parametrization protocol.
\end{abstract}

\maketitle

\section{Introduction}

In the atomic-scale modelling of (electro-)chemical reactions on surfaces, clusters or molecules it becomes increasingly apparent that solvent effects can have a qualitative influence on predicted reaction pathways and rates.\cite{Koper2013,Li2010} This poses a problem especially in first-principles electronic structure approaches where the large number of solvent molecules necessary to accurately represent bulk solvent properties render explicit solvation computationally impractical. This situation is further aggravated when considering non-negligible salt concentrations in the solvent. These are known to dramatically influence the electrochemistry,\cite{Israelachvili2011, Hunter2001} but demand even larger simulation boxes for realistic ionic strengths and correct thermodynamic sampling.

An approximate way to overcome this hurdle is to formally integrate out solvent degrees of freedom in order to treat the solvent outside of a so-called solvation cavity on the level of a dielectric continuum. While there are many possible choices of such mean-field, implicit solvation models\cite{Tomasi2005}, regarding the description of ionic effects Poisson-Boltzmann (PB) theory\cite{Gouy1910,Gouy1917,Chapman1913,Debye1923} has been remarkably successful\cite{Chen1997,Weetman1997,Fogolari2002,Hunter2001} and represents a wide-spread standard.
PB theory also treats the solvated ions on the level of a mean-field potential.\cite{Borukhov2000} In its original formulation it thereby considers only the electrostatic interactions between point-like ions in a dielectric medium to arrive at analytic expressions of appealing simplicity. At this level, PB theory has found application in many fields of molecular modelling, ranging from colloid science \cite{Verwey1948,Alexander1984} and polyelectrolytes\cite{Barrat1996} over surface science, electrochemistry and electrokinetics\cite{Hunter2001,Lyklem1995} to the simulation of biological systems\cite{Harries1998,Borukhov1998,Andelman1995}. 

Over the years there have been a number of approaches to improve upon the obvious shortcomings of such plain PB theory. These shortcomings comprise e.g.~the neglect of short-range steric repulsions between finite-sized ions, the general neglect of ionic correlations and fluctuations beyond the mean-field level and the neglect of dispersive contributions to the interactions. Corresponding attempts to improve on these aspects include liquid state theory approaches\cite{Kjellander1985,Kjellander1988,Blum1992,Levine1981,Outhwaite1983}, field theory expansions\cite{Netz1999,Netz2000}, ion correlation corrections\cite{Lukatsky1999, Stevens1990}, improvements on the description of solvent molecules\cite{Blum1992,Torrie1991,Otto1999} and finite-size correction for the ions\cite{Borukhov2000,Borukhov1997,Boschitsch2012,Chaudhry2011,Kraljlglic1996,Zhou2011,Abrashkin2007,Chu2007,Tresset2008} among a wide variety of other computational simulation approaches\cite{Guldbrand1984,Kjellander1992,Greberg1997}. In this work we focus on one of the most famous of such adaptations of PB theory, the size-modified PB (MPB) approach.\cite{Borukhov1997,Borukhov2000,Kraljlglic1996} While maintaining the mathematical and conceptual simplicity of the original PB formulation, MPB corrects specifically for finite ion sizes. For this, MPB theory makes use of a local excess free energy functional of the ion density\cite{Antypov2005}, which is based on a lattice gas model and corrects for steric ion-ion and ion-solvent repulsions via a blocking of lattice sites.\cite{Borukhov2000} MPB has proven to yield particularly good results for equally-sized ions and counter-ions with low charges, and is under active development to this day.\cite{Li2009,Zhou2011,Chu2007,Tresset2008,Boschitsch2012} 

We here present a new methodological approach to solving the MPB equation (MPBE) that is general enough to be applied to any kind of system. While in principle not tied to any particular electronic structure method, we focus on the coupling to density-functional theory (DFT). Such a coupling to DFT has recently been achieved for several other implicit solvation models. \cite{Fattebert2002,Fattebert2003,Scherlis2006,Andreussi2012,Kelly2005,Tomasi2005} Yet, previous attempts at specifically coupling DFT and MPB theory either relied on more approximate variants of the theory\cite{Letchworth-Weaver2012,Tannor1994} or were limited to specific solvation cavity geometries, such as planar surfaces\cite{Fang2013,Otani2006}.  More general schemes for arbitrary cavity shapes\cite{Jinnouchi2008,Fisicaro2016,Mathew2016} or approaches that can for instance exploit irregular integration grids are much less common and subject of current research. Precisely such irregular integration grids are essential, though, in resolving Coulomb singularities and orbital cusps in all-electron full-potential formulations of DFT using localized basis sets. Our new method is thus specifically geared towards solving the MPBE in such circumstances by formulating a Newton scheme in function space. Utilizing the properties of Green's functions (and multipole expansions) for the iterative solution of the linear subproblems then yields a self-consistent solvation free energy for arbitrary cavity shapes. Our approach furthermore includes a model for the well-known Stern layer which separates the diffusing ions from the solvation cavity by introducing non-mean-field ion-solute interactions.

This paper is organized as follows: After briefly introducing to MPB theory and deriving a Stern layer corrected version of the MPBE, we illustrate the efficient implementation specifically for the numeric-atomic orbital (NAO) based DFT framework FHI-aims \cite{Blum2009,Ren2012}. The fast convergence and high accuracy of the electrostatic potential, ion densities and solvation free energies in electrolyte solutions are then demonstrated for a range of neutral, organic molecules taken from the test set introduced by Shivakumar {\em et al.}\cite{Shivakumar2010}. In order to highlight the capabilities of our methodology we finally evaluate the concentration-dependent salt activity coefficient of KCl aqueous solutions and illustrate that the sensitivity of this quantity on the MPB ion parameters offers a novel route to systematically determine these parameters from tabulated activity coefficients.

\section{Theory}

\subsection{Poisson-Boltzmann Theory}

For the sake of self-containment and a consistent notation we first provide a brief outline of PB and MPB theory. In deriving the MPBE and the associated free energy functional we closely followed the work of Borukhov \textit{et al.}\cite{Borukhov2000} and Tresset\cite{Tresset2008}. Considered is the purely electrostatic interaction of a $z:z$ continuum electrolyte with a solute, which could e.g.~be one or more molecules, a cluster or a solid surface. The $z:z$ electrolyte of bulk dielectric permittivity $\varepsilon^{\rm s,bulk}$ hereby contains cations and anions of equal opposite charge, $z = z_+ = -z_-$, and equal bulk ion concentrations, $c^{\rm s,bulk} = c^{\rm s,bulk}_+ = c^{\rm s,bulk}_-$. The total charge distribution of the solute derives from electronic and nuclear contributions
\begin{equation}
n_{\rm sol}(\bm{r}) = n_{\rm el}(\bm{r}) \underbrace{- \sum_\mathrm{at} Z_\mathrm{at}\delta(\bm{r}-\bm{R}_\mathrm{at})}_{n_\mathrm{nuc}} \quad ,
\end{equation}
where the sum ranges over all nuclear centers ``at'' of charge $Z_\mathrm{at}$ and located at positions $\lbrace \bm{R}_\mathrm{at}\rbrace$. Throughout this work we will use atomic units and the typical sign convention known from DFT, so electronic charge densities have a positive sign. This charge distribution gives rise to an electric field $\bm{E}$, which causes a displacement field $\bm{D}$ in the surrounding dielectric medium. In isotropic media this field can be approximated as
\begin{equation}
\bm{D}(\bm{r}) = \varepsilon(\bm{r})\bm{E}(\bm{r}) \quad,
\end{equation}
where $\varepsilon(\bm{r})$ is the position-resolved, static dielectric response function of the medium\cite{Fattebert2002}, i.e. a generalization of the macroscopic concept of dielectric permittivity, which for instance is 1 in vacuum and $\approx 80$ (78.36 in this work) in bulk water. In this formulation, all non-locality in the solvent response is neglected, which is valid for small enough electric fields causing only a slight variation of the displacement fields.\cite{Bonthuis2013}

\begin{figure}[htb]
\centering
\includegraphics[width=1.\columnwidth]{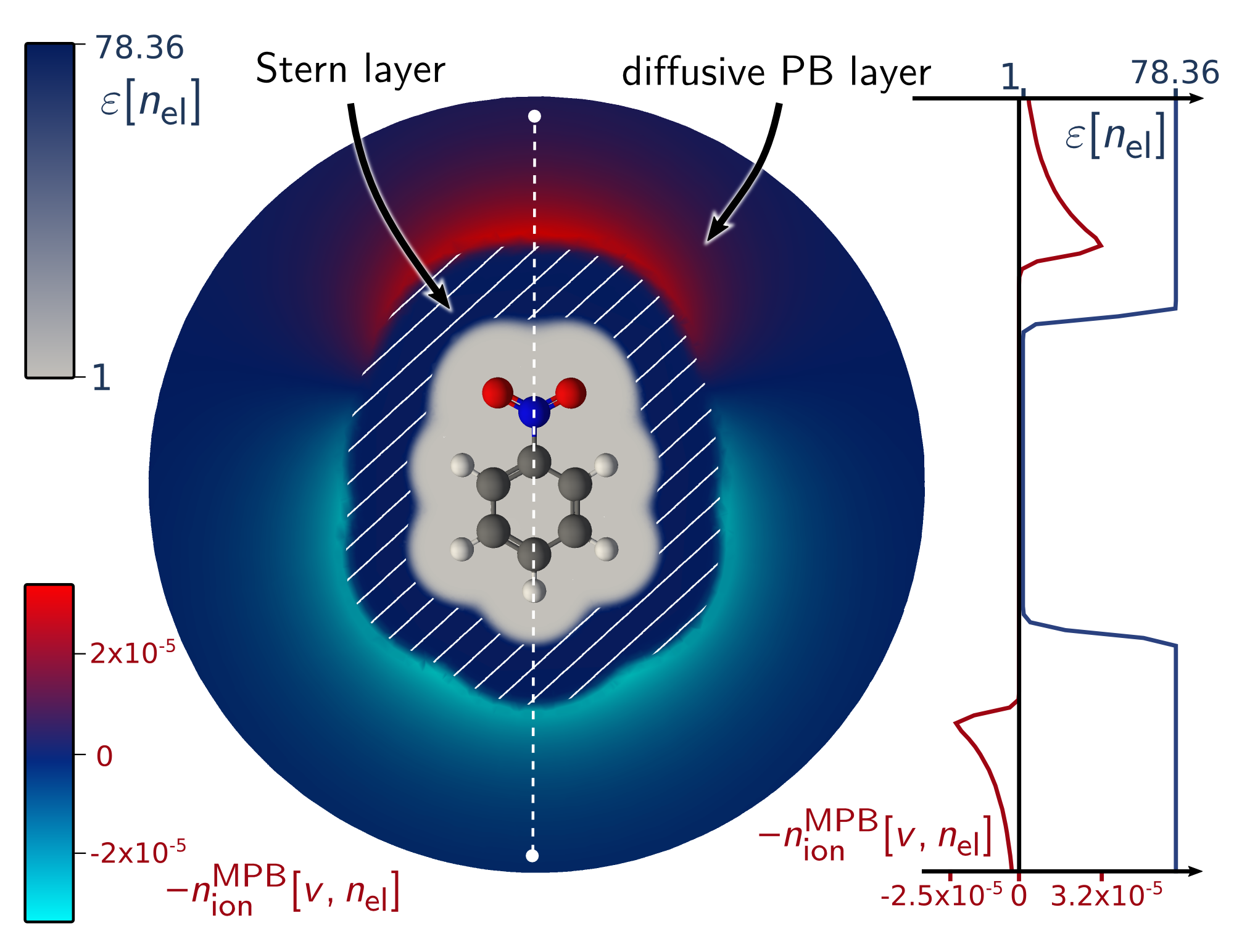}
\caption{Left panel: Schematic visualization of nitrobenzene dissolved in water containing a 1M 1:1 electrolyte as described at the level of MPB theory with additional ion exclusion (Stern) layer. The gray innermost part around the molecule denotes the solvation cavity, in which $\varepsilon[n_{\rm el}(\bm{r})] = 1$, the remaining area denotes the solvent region, in which $\varepsilon[n_{\rm el}(\bm{r})] = \varepsilon^\mathrm{s,bulk}$. A contour plot of the ionic charge density $n_\mathrm{ion}^\mathrm{MPB}[v(\bm{r}),n_{\rm el}(\bm{r})]$ is overlayed (using the intuitive sign convention), where the striped region depicts the ion-free Stern layer. Right panel: $\varepsilon[n_{\rm el}(\bm{r})]$ and $
n_\mathrm{ion}^\mathrm{MPB}[v(\bm{r}),n_{\rm el}(\bm{r})]$ along the dashed line through the center of the molecule.}
\label{fig:system_scheme}
\end{figure}

The dielectric medium is excluded from a volume around the solute -- the solvation cavity -- and assumed to smoothly adapt bulk properties at increasing distances from the solute. As schematically illustrated in Fig.~\ref{fig:system_scheme} this therefore implies $\varepsilon(\bm{r}) = 1$ within the cavity and ${\varepsilon(\bm{r}) = \varepsilon^{\rm s,bulk}}$ far away. For the transition in between we specifically use a parametrization in terms of the electron density $n_{\rm el}$, as it is the finite electron density tails leaking outside of the cavity that govern the near-solute dielectric properties in this local formulation. As a monotonous function of the electron density, which in turn implicitly depends on the electrostatic potential $v(\bm{r})$, $\varepsilon(\bm{r})$ then couples $\bm{D}(\bm{r})$ non-linearly to $\bm{E}(\bm{r})$ at any point in space. The connection between the displacement field and the charge distribution of the solute is given by the Poisson equation
\begin{equation}
-\nabla \bm{D}(\bm{r}) \;=\; \nabla \cdot \left[\varepsilon[n_{\rm el}(\bm{r})]\nabla v(\bm{r})\right] \;=\; -4\pi n_{\rm sol}(\bm{r}) \quad.
\label{eq:Poisson}
\end{equation}
In PB theory the hitherto unaccounted salt ions are introduced by simply adding a continuous ionic charge density $n_\mathrm{ion}^\mathrm{PB}(\bm{r})$ to the source term of this Poisson equation. This leads to the Poisson-Boltzmann equation
\begin{equation}
\nabla \cdot \left[\varepsilon[n_{\rm el}(\bm{r})]\nabla v(\bm{r})\right] \;=\; -4\pi n_{\rm sol}(\bm{r}) - 4\pi n_\mathrm{ion}^\mathrm{PB}(\bm{r}) \quad ,
\label{eq:PBE}
\end{equation}
with
\begin{equation}
n_\mathrm{ion}^\mathrm{PB}(\bm{r}) \;=\; z \left[ c^{\rm s}_+(\bm{r}) - c^{\rm s}_-(\bm{r}) \right] \quad ,
\label{eq:PBE_fv}
\end{equation}
where $c^{\rm s}_+(\bm{r})$ and $c^{\rm s}_-(\bm{r})$ are the spatially-dependent concentrations of the dissolved cations and anions, respectively. 

As these ions are mobile and also subject to Coulomb interactions, their distributions depend generally on the overall electrostatic potential. In a mean-field picture, this dependence would be rigorously captured by the potential of mean force that would appropriately average over the ionic fluctuations.\cite{Fogolari2002,Attard1996,Bonthuis2013} In PB theory this dependence is instead approximately accounted for through the mean-field electrostatic potential, i.e. ${c^{\rm s}_+(\bm{r}) = c^{\rm s}_+[v(\bm{r})]}$, ${c^{\rm s}_-(\bm{r}) = c^{\rm s}_-[v(\bm{r})]}$ and correspondingly also ${n_\mathrm{ion}^\mathrm{PB}(\bm{r}) = n_\mathrm{ion}^\mathrm{PB}[v(\bm{r})]}$. Expressions for these dependencies are then e.g.~derived from statistical models parameterizing partition functions and minimizing the resulting free energy expressions with respect to the electrostatic potential (see below).\cite{Borukhov2000}

\subsection{Size-Modified Poisson-\\Boltzmann theory including Stern-layer correction}

The various existing flavors of PB theory differ mainly in the form considered for the ionic concentrations $c^{\rm s}_\pm[v(\bm{r})]$ and their dependence on $v(\bm{r})$.\cite{Grochowski2008} In the original formulation arising from Gouy-Chapman\cite{Gouy1910,Chapman1913} or Debye-H\"uckel theory\cite{Debye1923} point-like and Boltzmann distributed ions were assumed. One severe disadvantage of this approach, as pointed out by a number of authors\cite{Borukhov2000,Grochowski2008} and proven by experiments\cite{Cervera2010,Kosmulski1995}, is the disregard of the finite ion sizes which leads even for small ions to an overestimation of the ionic charge accumulation close to the solute and therefore to overall erroneous ionic effects, especially for charged solutes creating high electrostatic potentials $|v(\bm{r})|$ at the cavity surface.

MPB theory, as e.g.~derived by Borukhov {\em et al.} \cite{Borukhov2000}, aims to correct for this by giving the ions an explicit size. Its construction is based on a lattice gas model where solvent molecules and ions compete for empty lattice sites of a cell size $a$, where each lattice site can only be occupied by one particle at a time. From the partition function of this model system one derives a correction to the free energy functional due to the entropy of the electrolytic system, which depends strongly on the size of the lattice cells and the temperature. Minimizing this functional with respect to the electrostatic potential yields a modified ionic charge density $n_\mathrm{ion}^\mathrm{MPB}(\bm{r})$. Including an additional Stern-layer correction ({\em vide infra}) this density reads
\begin{align}
n_\mathrm{ion}&^\mathrm{MPB}[v(\bm{r}),n_{\rm el}(\bm{r})] \;=\; -2 z c^\mathrm{s,bulk} \alpha_{\rm ion}[n_{\rm el}(\bm{r})] \nonumber\\&\cdot\frac{\mathrm{sinh}(z \beta v(\bm{r}))}{1-\phi_0+\phi_0 \alpha_{\rm ion}[n_{\rm el}(\bm{r})]\mathrm{cosh}\left(z \beta v(\bm{r})\right)} \quad .
\label{eq:MPBE}
\end{align} 
Here $\beta = 1/k_\mathrm{B}T$ is the inverse temperature, where $k_\mathrm{B}$ is the Boltzmann constant, $T$ the temperature, and $\phi_0 = 2 a^3 c^\mathrm{s,bulk}$ denotes the volume fraction of lattice cells occupied by ions. As also illustrated by the example in Fig.~\ref{fig:system_scheme} and in contrast to standard PB theory (which is recovered for $a = 0$), this ion distribution function now remains bounded even in the limit of very large local electrostatic potentials $|v(\bm{r})|$. This comes at the expense of an additional parameter $a$ that needs to be determined in the application to real systems, cf. Section~\ref{sec:activ}.

Equation (\ref{eq:MPBE}) additionally contains a so-called ion exclusion function $\alpha_{\rm ion}(\bm{r})=\alpha_{\rm ion}[n_{\rm el}(\bm{r})]$ which is not present in the original MPB formulation of Borukhov {\em et al.} \cite{Borukhov2000}. This function corrects for the shortcoming that in the original MPB theory ions can in principle still approach the solute atoms as close as half of the lattice cell length $a$, and could therefore even lie deep within the electron distribution $n_{\rm el}(\bm{r})$ of the solute. In reality, however, the diffusive ion layer lies outside the solvation cavity and is separated from this by an exclusion zone, often called the Stern layer \cite{Stern1924}. In this layer, solvent molecules are oriented around the charge and sometimes (in the case of overall charged solutes) also ions are directly adsorbed. Yet, most importantly, no diffusive ion distribution exists. The origin of this layer is partly attributed to ion-charge distribution non-mean-field interactions, as e.g.~dispersion or short-range repulsion interactions\cite{Bostrom2005}, which are not included in the original MPB formulation. In explicitly correcting for this Stern layer, we thereby follow a reasoning analogous to Jinnouchi \textit{et al.}\cite{Jinnouchi2008}, who included a short-range ion-solute repulsion operator in the Hamiltonian which Dabo \textit{et al.}\cite{Dabo2008} expressed as an ion exclusion function. We note that this stands in contrast to other authors like Otani \textit{et al.}\cite{Otani2006} who claim the original MPBE to be sufficient to treat size effects or Bostr\"om {\em et al.}, who apply an additional repulsion/dispersion operator but use the standard PBE instead of the MPBE\cite{Bostrom2005}. 

For a maximum of generality our implementation allows the use of finite-sized ions, while still retaining the effective correction for the Stern layer (cf. also Harris \textit{et al.}\cite{Harris2014}). The resulting ion exclusion function $\alpha_{\rm ion}[n_{\rm el}(\bm{r})]$ then restricts the region of finite entropy of the electrolyte to the diffusive part of the system, where both solvent molecules and ions compete for lattice sites. We point out that similar treatments exist in literature, but are often based on sharp step functions.\cite{Fang2013} In contrast and as further discussed in Section~\ref{sec:methods-dielec}, our exclusion function is set to $\alpha_{\rm ion}=0$ close to the solute and $\alpha_{\rm ion}=1$ far away, but with a smoothly modeled transition in between. Similar to our dielectric model, we choose a parameterization of $\alpha_{\rm ion}[n_{\rm el}(\bm{r})]$ in terms of the electron density in order to capture the repulsion from the solute charge reaching beyond the cavity. 

Regardless of the actual treatment of the ions, both PBE and MPBE are non-linear partial differential equations. As such, their structure precludes an analytic solution for all but the simplest solutes, while at the same time making them difficult to solve numerically.\cite{Holst1993} However, for weak solute potentials -- e.g.~for neutral solutes -- the equations can often be linearized to obtain the famous linearized PBE (LPBE) known mainly from Debye-H\"uckel theory\cite{Debye1923}. For our case of the MPBE, the linearized ionic charge density in eq.~(\ref{eq:PBE_fv}) is given by
\begin{equation}
n_\mathrm{ion}^\mathrm{LPB}[v(\bm{r}),n_{\rm el}(\bm{r})] = -\frac{1}{4\pi}\bar{\kappa}^2[n_{\rm el}(\bm{r})]v(\bm{r}) \quad ,
\label{eq:LPBEcharge}
\end{equation}
which correspondingly leads to a LPBE that is linear in $v$
\begin{align}
\hat{L}_0v(\bm{r})&= \biggl(\nabla \cdot \left[\varepsilon[n_{\rm el}(\bm{r})] \nabla\right] - \bar{\kappa}^2[n_{\rm el}(\bm{r})] \biggr) v(\bm{r}) \nonumber\\&= -4\pi n_{\rm sol}(\bm{r}) \quad .
\label{eq:LPBE}
\end{align}
Here, 
\begin{equation}
\bar{\kappa}^2[n_{\rm el}(\bm{r})] = \frac{\alpha_{\rm ion}[n_{\rm el}(\bm{r})]}{1+\phi_0(\alpha_{\rm ion}[n_{\rm el}(\bm{r})]-1)}\varepsilon^\mathrm{s,bulk} \kappa^2 \quad ,
\end{equation}
with the Debye-H\"uckel parameter
\begin{equation}
\kappa = \sqrt{\frac{8\pi c^\mathrm{s,bulk} z^2 \beta}{\varepsilon^\mathrm{s,bulk}}} \quad.
\end{equation}
Equations of the LPBE type are indeed much more convenient to solve, but due to the limitations on the magnitude of the electrostatic potential, such as e.g.~$|v(\bm{r})| \ll 25$\,mV for monovalent ions at room temperature\cite{Kilic2007a,Borukhov1997,Debye1923}, not universally applicable. Nevertheless, in Section~\ref{sec:methods} below we will demonstrate how the solution of LPB type of equations can be used as a stepping stone to solving the fully non-linear MPBE.

\subsection{Free Energy Functional}
\label{sec:fef}

Next to the solvent's influence on the electronic structure of the solute, the main observable of interest in an implicit solvation scheme is the solvation (or electrolyzation) free energy $\Delta G_\mathrm{sol}$. It is defined as 
\begin{align}
\Delta G_\mathrm{sol}&= \Omega_\circ(\varepsilon^{\rm s,bulk},c^\mathrm{s,bulk}, n_{\rm sol}(\bm{r})) \nonumber\\&- \Omega_\circ(\varepsilon^{\rm s,bulk}=1, c^\mathrm{s,bulk}=0,n_{\rm sol}(\bm{r})) \nonumber\\&- \Omega_\circ(\varepsilon^{\rm s,bulk},c^\mathrm{s,bulk}, n_{\rm sol}(\bm{r})=0) \quad , 
\label{eq:solvenergyeq}
\end{align}
where ${\Omega_\circ(\varepsilon^{\rm s,bulk},c^\mathrm{s,bulk}, n_{\rm sol}(\bm{r}))}$ is the free energy of the solute embedded into a solvent connected to reservoir of solvent molecules and ions, ${\Omega_\circ(\varepsilon^{\rm s,bulk}=1, c^\mathrm{s,bulk}=0,n_{\rm sol}(\bm{r}))}$ the free energy of the solute in vacuum, and ${\Omega_\circ(\varepsilon^{\rm s,bulk},c^\mathrm{s,bulk}, n_{\rm sol}(\bm{r})=0)}$ the free energy of the pure electrolyte. Another similar core quantity is the ion effect on the solvation free energy, defined as
\begin{equation}
\Delta\Delta G_\mathrm{ion} = \Delta G_\mathrm{sol}(c^\mathrm{s,bulk}) - \Delta G_\mathrm{sol}(c^\mathrm{s,bulk} = 0) \quad .
\label{eq:ionenergyeq}
\end{equation} 

In MPB theory the required free energy expressions are obtained by minimizing a phenomenologically or field-theoretically derived grand potential $\Omega^\mathrm{mf}_{\varepsilon,\alpha_\mathrm{ion}}[v(\bm{r}),c^{\rm s}_+(\bm{r}), c^{\rm s}_-(\bm{r})]$ with respect to the potential and ion distributions.\cite{Borukhov2000} Since this only accounts for the mean-field electrostatic interactions within the solvent and between solute and solvent, additional terms need to be considered in an extended free energy functional when aiming to use MPB theory in combination with a first-principles electronic structure approach like DFT. Specifically we therefore consider the following free energy functional and from now on drop all $\bm{r}$-dependencies in the equations for improved legibility

\begin{align}
\nonumber
\Omega&_{\varepsilon,\alpha_{\rm ion}}[v,n_{\rm el},c^\mathrm{s}_+,c^\mathrm{s}_-]  =
T^\mathrm{S}[n_{\rm el}] + E^\mathrm{xc}[n_{\rm el}] \\ 
& + \underbrace{\Omega^\mathrm{mf}_\varepsilon[v,n_{\rm el},c^\mathrm{s}_+,c^\mathrm{s}_-] + \Omega^\mathrm{mf}_{\alpha_{\rm ion}}[n_{\rm el},c^\mathrm{s}_+,c^\mathrm{s}_-]}_{\Omega^\mathrm{mf}_{\varepsilon,\alpha_{\rm ion}}[v,n_{\rm el},c^\mathrm{s}_+,c^\mathrm{s}_-]} \;\nonumber\\&+\; \Omega^\mathrm{non-mf}[n_{\rm el}] \quad.
\label{eq:dft21}
\end{align}
Here, $T^\mathrm{S}[n_{\rm el}]$ and $E^\mathrm{xc}[n_{\rm el}]$ are the usual kinetic energy and exchange-correlation (xc) energy functionals of DFT accounting for corresponding contributions from the solute electrons, respectively. No separate Hartree energy functional is considered. Instead, the mean-field electronic interactions within the solute are appropriately included in $\Omega^\mathrm{mf}_\varepsilon[v,n_{\rm el},c^\mathrm{s}_+,c^\mathrm{s}_-]$ which thus accounts for all corresponding interactions in solute, electrolyte and between solute and electrolyte. The Stern layer correction gives rise to an additional mean-field term $\Omega^\mathrm{mf}_{\alpha_{\rm ion}}[n_{\rm el},c^\mathrm{s}_+,c^\mathrm{s}_-]$ (see below) which we treat jointly with the regular MPB term as $\Omega^\mathrm{mf}_{\varepsilon,\alpha_{\rm ion}}[v,n_{\rm el},c^\mathrm{s}_+,c^\mathrm{s}_-]$. Finally, there are also non-mean-field, non-electrostatic interactions between solute and solvent $\Omega^\mathrm{non-mf}[n_{\rm el}]$.\cite{Scherlis2006,Andreussi2012} While in principle far from trivial, Andreussi \textit{et al.} have shown that solvation free energy contributions arising from $\Omega^\mathrm{non-mf}[n_{\rm el}]$ can to a good degree of accuracy be modeled as a linear function of volume and surface of the cavity formed around the solute.\cite{Andreussi2012} As further described in Section~\ref{sec:methods-dielec} below, both can be written as integrals over the solute charge density, which is why we consider here only a dependence on $n_{\rm el}$ for this term. While eq.~(\ref{eq:dft21}) thus contains a number of interactions beyond regular MPB theory, we stress that further interactions could still be included. This concerns notably an account for the entropies of the solute electrons and nuclei at finite temperatures, or entropic terms arising from the kinetic energies of solvent molecules and ions \cite{Tresset2008}. For the targeted free energy differences ($\Delta G_{\rm sol}$, $\Delta \Delta G_{\rm ion}$) these terms are not expected to contribute significantly, which is why they are neglected here in accordance with most implicit solvation methods.\cite{Andreussi2012,Scherlis2006,Fattebert2002}

The resulting extended free energy functional $\Omega_{\varepsilon,\alpha_{\rm ion}}[v,n_{\rm el},c^\mathrm{s}_+,c^\mathrm{s}_-]$ depends on the electron density, the electrostatic potential and the ion distributions, and needs to be minimized with respect to all of them to yield the required free energy expression. The minimization with respect to the ion distributions concerns only the $\Omega^\mathrm{mf}_{\varepsilon,\alpha_{\rm ion}}[v,n_{\rm el},c^\mathrm{s}_+,c^\mathrm{s}_-]$-term and can be done separately. To describe the energy shift due to the additional Stern layer ion-solute interaction we relate the ion exclusion function $\alpha_{\rm ion}$ to a repulsion potential via ${\alpha_{\rm ion}[n_{\rm el}] = \exp(-\beta v^\mathrm{rep})}$ and then obtain a free energy functional contribution analogous to the work of Jinnouchi and co-workers\cite{Jinnouchi2008,Jinnouchi2011}. While the physics is thus the same, our motivation to use an exclusion function instead of directly imposing a repulsion potential is thereby mainly based on numerical efficiency arguments. The resulting expression then reads
\begin{align}
&\Omega^\mathrm{mf}_{\alpha_{\rm ion}}[n_{\rm el},c^{\rm s}_+,c^{\rm s}_-] = \int \mathrm{d}\bm{r}(c^{\rm s}_+ + c^{\rm s}_-)v^\mathrm{rep} \nonumber\\&= -\frac{1}{\beta}\int \mathrm{d}\bm{r}(c^{\rm s}_+ + c^{\rm s}_-)\ln(\alpha_{\rm ion}[n_{\rm el}]) \quad .
\label{eq:phenoFreeEn}
\end{align}
Minimizing $\Omega^\mathrm{mf}_{\varepsilon,\alpha_{\rm ion}} = \Omega^\mathrm{mf}_\varepsilon + \Omega^\mathrm{mf}_{\alpha_{\rm ion}}$ with respect to the $c^{\rm s}_{\pm}$'s yields a closed expression for the equilibrium distribution of ions around the solute corresponding to eq.~(\ref{eq:MPBE}), cf. supplementary information (SI). Inserting this into the functional leads to
\begin{align}
\Omega^\mathrm{mf}&_{\varepsilon,\alpha_{\rm ion}}[v,n_{\rm el}]\nonumber\\&= \int \mathrm{d}\bm{r}\biggl\{-\frac{ \varepsilon[n_{\rm el}]}{8\pi}|\nabla v|^2 + n_{\rm sol}v\nonumber\\&-\frac{1}{\beta a^3} \ln\left(1+\frac{\phi_0}{1-\phi_0}\alpha_{\rm ion}[n_{\rm el}]\cosh(\beta z v)\right)\biggr\} \quad ,
\label{eq:MPBEFreeEnFunctional}
\end{align}
with a full derivation provided in the SI.

\subsection{Modified Kohn-Sham equations and minimum free energy expression}

Using eq.~(\ref{eq:MPBEFreeEnFunctional}) the extended free energy functional $\Omega_{\varepsilon,\alpha_{\rm ion}}$ of eq.~(\ref{eq:dft21}) remains a functional of the electron density and the electrostatic potential. Minimization with respect to the latter yields the MPBE of eqs.~(\ref{eq:PBE}) and (\ref{eq:MPBE}). Minimization with respect to the former will -- in full analogy to standard Kohn-Sham DFT \cite{Kohn1965} -- lead to modified Kohn-Sham equations that additionally account for the solvent and ion effects on the solute electronic structure. Both MPBE and modified Kohn-Sham equations then have to be solved self-consistently to obtain the ground-state electron density, the electrostatic potential and subsequently the ground-state free energy.

In order to fulfill the orthonormality constraint of the single-electron Kohn-Sham states minimization with respect to the electron density 
proceeds in practice via the Lagrangian
\begin{align}
\mathcal{L}&[v,n_{\rm el}] = \Omega_{\varepsilon,\alpha_{\rm ion}}[v,n_{\rm el}] \nonumber\\&+ \sum_{l=1}^{N}\sum_{k=1}^{N} \lambda_{lk} \left[\int \mathrm{d}\bm{r}\psi_l^* \psi_k - \delta_{lk} \right] \quad ,
\label{eq:lagrange_dft}
\end{align}
with the Lagrange multipliers $\lambda_{lk}$, and $N$ being the total number of single-particle Kohn-Sham states $\psi_l$. The stationary state of this Lagrange functional with respect to all $\psi_l$ yields the electronic ground state. If $\mathcal{L}$ contained only the regular DFT energy functional, this minimization would yield the standard Kohn-Sham equations. Due to the additional solvation terms coming from $\Omega^\mathrm{mf}_{\varepsilon,\alpha_{\rm ion}}$ and $\Omega^\mathrm{non-mf}$ in $\Omega_{\varepsilon,\alpha_{\rm ion}}[v,n_{\rm el}]$ additional terms will arise in these equations. In the following we will thereby concentrate on the terms arising from $\Omega^\mathrm{mf}_{\varepsilon,\alpha_{\rm ion}}$. The effect of $\Omega^\mathrm{non-mf}$ will instead only be treated as a non-self-consistent post-correction of the free energy. The results of Section~\ref{sec:results-Esolv} suggest this to imply only a negligible error, while simultaneously allowing to avoid the demanding computation of second derivatives of the free energy functional and avoiding numerical instabilities as observed by several authors\cite{Steinmann2016}. 

With this simplification and as derived in the SI the only additional term to the single-particle Kohn-Sham Hamiltonian $\hat{h}^{\rm KS}$ arising from the electron density dependence of the dielectric function and the ion exclusion function is
\begin{align}
\delta& v^\mathrm{KS,MPB}_{\varepsilon,\alpha_{\rm ion}}
=-\frac{1}{8\pi}\frac{\partial \varepsilon[n_{\rm el}]}{\partial n_{\rm el}}|\nabla v|^2 \nonumber\\&-\frac{\phi_0}{\beta a^3} \frac{\partial \alpha_\mathrm{ion}[n_{\rm el}]}{\partial n_{\rm el}}\frac{  \cosh(\beta z v)}{1-\phi_0+\phi_0\alpha_\mathrm{ion}[n_{\rm el}]\cosh(\beta z v)} \quad .
\label{eq:extra_KSHam}
\end{align}
Self-consistent solution of the thus modified Kohn-Sham equations together with the MPBE yields the ground-state electron density $n_{{\rm el},\circ}$, the ground-state electrostatic and xc potential $v_{\circ}$ and $v^\mathrm{xc}_\circ$, respectively, the ground-state single-particle eigenvalues $\epsilon_{l,\circ}$ and the ground-state eigenstate occupation numbers $f_{l,\circ}$. Using this the free energy is finally obtained as (cf. supplementary information)
\begin{widetext}
\begin{align}
\Omega&_\circ(\varepsilon^{\rm s,bulk},c^\mathrm{s,bulk}, n_{\rm sol,\circ}) =\sum\limits_{l=1}^{N_\mathrm{states}}f_{l,\circ} \epsilon_{l,\circ} - 
\displaystyle\int\mathrm{d}\bm{r} n_{{\rm el},\circ} v^\mathrm{xc}_\circ +  E^\mathrm{xc}[n_{{\rm el},\circ}] -
\frac{1}{2}\displaystyle\int\mathrm{d}\bm{r} n_{{\rm el},\circ} v_\circ + \frac{1}{2}\int \mathrm{d}\bm{r} n_\mathrm{nuc} v_\circ\nonumber\\
&-\displaystyle\int\mathrm{d}\bm{r} n_{{\rm el},\circ} \delta v^\mathrm{KS,MPB}_{\varepsilon,\alpha_{\rm ion}}+\int\mathrm{d}\bm{r}\biggl\{-\frac{1}{2}n_\mathrm{ion}^\mathrm{MPB}[v_{\circ},n_\mathrm{{\rm el},\circ}]v_{\circ} - \frac{1}{\beta a^3}\ln\left(1+\frac{\phi_0}{1-\phi_0}\alpha_{\rm ion}[n_{{\rm el},\circ}]\cosh(\beta z v_{\circ})\right)\biggr\} \nonumber\\
&+\Omega^\mathrm{non-mf}[n_\mathrm{{\rm el},\circ}] \quad,
\label{eq:energymin4}
\end{align}
\end{widetext}
with $N_\mathrm{states}$ being the number of possibly degenerated single-particle Kohn-Sham states $\psi_l$. The first line contains the regular energy contributions known from DFT, including the exchange-correlation energy $E^{\rm xc}$, the nuclear-nuclear repulsion energy and double counting correction. The latter two are thereby represented by the last two terms by adding and substracting the interaction of the electrons with the external potential to increase the stability of the energy expression.\cite{Blum2009} The second line contains the additional solvation term arising within the Kohn-Sham framework, the third and fourth lines are the free energy contributions from the MPB grand potential $\Omega^\mathrm{mf}_{\varepsilon,\alpha_{\rm ion}}$ and the last line the non-mean-field contribution $\Omega^\mathrm{non-mf}$ that we here only treat as a post-correction. The regular DFT contribution is thereby already written in a way that allows straightforward integration into a full-potential DFT code like FHI-aims\cite{Blum2009}. 

\subsection{Cavity definition and non-mean-field contributions}
\label{sec:methods-dielec}

For its application, the derived MPB solvation model requires functional forms for the dielectric and ion exclusion functions, $\varepsilon[n_{\rm el}]$ and $\alpha_{\rm ion}[n_{\rm el}]$, respectively. The transition of both functions from the bulk solution value to the value inside the solute defines the solvation cavity: $\varepsilon[n_{\rm el}]$ for the solvent and $\alpha_{\rm ion}[n_{\rm el}]$ the additional Stern layer for the ions. In principle, the corresponding transition of both physical counterparts, the spatially resolved permittivity and the ion exclusion function, respectively, are highly solute and solvent dependent. The choice of smooth switching functions depending solely on the solute electron density has nevertheless generally shown satisfactory transferability in the prediction of solvation free energies.\cite{Andreussi2012,Fattebert2002,Scherlis2006} For the present first proof-of-concept of our solvation model we therefore simply employ a functional form developed originally by Andreussi {\em et al.}.\cite{Andreussi2012} Adaptation of this functional form in the implementation in FHI-aims is trivial and future work will concentrate on a systematic evaluation and parametrization of this form in dedicated applications.

The functional form employed for the dielectric function is a smoothed step function of the form\cite{Andreussi2012}
\begin{equation}
\varepsilon_{n_\mathrm{min},n_\mathrm{max}}[n_{\rm el}] = 
\begin{cases}
1 & n_{\rm el} > n_\mathrm{max},\\
\mathrm{e}^{t\left(\ln(n_{\rm el})\right)} & n_\mathrm{min} < n_{\rm el} < n_\mathrm{max} \quad ,\\
\varepsilon^\mathrm{s,bulk} & n_{\rm el} < n_\mathrm{min}\\
\end{cases}
\label{dielectric_function}
\end{equation}
with
\begin{align}
t\left(\ln(n_{\rm el})\right) = &\frac{\ln(\varepsilon^\mathrm{s,bulk})}{2\pi}\biggl[2\pi
\frac{\ln(n_\mathrm{max})-\ln(n_{\rm el})}{\ln(n_\mathrm{max})-
\ln(n_\mathrm{min})}\nonumber\\&-\sin\left(2\pi\frac{\ln(n_\mathrm{max})-\ln(n_{\rm el})}{\ln(n_\mathrm{max})-\ln(n_\mathrm{min})}\right)\biggr] \quad .
\end{align}
This function goes to one for large $n_{\rm el}$ close to the nuclei and switches smoothly to $\varepsilon^\mathrm{s,bulk}$ for low $n_{\rm el}$ far away. The transition region -- i.e. its position and width with respect to the electron density -- is controlled by the two parameters $n_\mathrm{min}$ and $n_\mathrm{max}$. The main benefit of this particular functional form is that its gradients are exactly zero outside of the transition region, and also $\nabla \ln(\varepsilon[n_{\rm el}])$ that appears in our developed solution scheme for the MPBE (cf. eqs.~(\ref{eq:newton_1}) and (\ref{eq:newton_2}) below) is a smooth function.\cite{Andreussi2012} This increases the numerical stability and the convergence with respect to the multipole expansion performed in a NAO-based DFT code like FHI-aims ({\em vide infra}).

For simplicity the same functional form is also chosen for the ion exclusion function
\begin{align}
\alpha&_{\rm ion,n_\mathrm{min}^\alpha,n_\mathrm{max}^\alpha}[n_{\rm el}] \nonumber\\&=
\begin{cases} 
0 & n_{\rm el} > n^\alpha_\mathrm{max}\\
\frac{1}{\varepsilon^\mathrm{s,bulk}-1}(\mathrm{e}^{t\left(\ln(n_{\rm el})\right)}-1)
 & n^\alpha_\mathrm{min} < n_{\rm el} < n^\alpha_\mathrm{max} \quad .\\
1 & n_{\rm el} < n^\alpha_\mathrm{min}\\
\end{cases}
\end{align}
This function depends on separate boundary density parameters, $n^\alpha_\mathrm{min/max}$, since the ion exclusion does not {\em a priori} have to show the same density dependence as the permittivity\cite{Harris2014}. Physically, the two transitions should be related though, which is why we define these boundary density parameters through a shift $d_{\alpha_{\rm ion}}$ and a scaling parameter $\xi_{\alpha_{\rm ion}}$ with respect to the parameters of the dielectric transition
\begin{align}
n&^\alpha_{\rm min/max} \nonumber\\&= \exp\left({a_{\rm min/max} \pm (a_\mathrm{max} - a_\mathrm{min})\frac{1-\xi_{\alpha_{\rm ion}}}{2}}\right)
\end{align}
with
\begin{align}
a&_{\rm min/max} \nonumber\\&= \ln(n_{\rm min/max}) +  \left(\ln(n_\mathrm{min}) - \ln(n_\mathrm{max})\right)d_{\alpha_{\rm ion}} \quad .
\end{align}
$d_{\alpha_{\rm ion}} > 0$ then corresponds to the inclusion of a Stern layer or non-diffusive region around the solvation cavity and a lowering (raise) of $\xi_{\alpha_{\rm ion}}$ to a sharpening (smoothening) of the Stern layer transition.

Finally, the choice of the electron density used in the definition of both transition functions deserves further mention. In principle, the evaluation could either be based on the true electron density $n_\mathrm{el}$ in each self-consistent field (SCF) step of the DFT solver, or it could be based on a rigid electron density obtained by mere superposition of free atom densities. In contrast to other authors \cite{Fang2013} we hitherto found only a negligible impact of a fully self-consistent cavity on the SCF convergence as long as the cavity lies within reasonable distances to the charge distribution, cf.~Fig.~\ref{fig:nitro_scf} below. All calculations in this work are correspondingly performed using the self-consistent density, through which we are able to model the mutual influence of the dielectric function and the electron density.

The solvation cavity defined through the transition functions also governs the non-mean-field part of the free energy functional $\Omega^\mathrm{non-mf}[n_{{\rm el},\circ}]$. In the differences of eq.~(\ref{eq:solvenergyeq}) defining $\Delta G_{\rm sol}$, this non-mean-field part gives rise to a free energy contribution due to the exclusion of solvent molecules from the cavity and non-bonded short-range, as well as dispersion interactions
\begin{equation}
\Delta G^\mathrm{non-mf}_{\rm sol} = \Delta G^\mathrm{cav} + \Delta G^\mathrm{rep} + \Delta G^\mathrm{dis} \quad ,
\end{equation}
respectively. In this work we employ the effective parametrization for these terms suggested by Andreussi \textit{et al.}\cite{Andreussi2012}, which provide these terms as mere functions of the ``quantum surface'' $S$ and the ``quantum volume'' $V$ of the solvation cavity
\begin{equation}
\Delta G^\mathrm{non-mf}_{\rm sol} = \left(\alpha+\gamma\right) S + \beta V \quad,
\end{equation}
with $\gamma$ the surface tension of the solvent. $\alpha$ and $\beta$ constitute additional free parameters of the model. $V$ and $S$ are hereby defined as
\begin{equation}
V=\int\mathrm{d}\bm{r}\vartheta[n_{\rm el}]
\end{equation}
and 
\begin{align}
S&=\int \mathrm{d}\bm{r}\biggl\{\left(\vartheta\left[n_{\rm el}-\frac{\Delta}{2}\right]-\vartheta\left[n_{\rm el}+\frac{\Delta}{2}\right]\right)\nonumber\\&\times\frac{|\nabla n_{\rm el}|}{\Delta}\biggr\} \quad,
\label{quantum_surface}
\end{align}
with the switching function $\vartheta$ defined in terms of the chosen dielectric function
\begin{equation}
\vartheta\left[n_{\rm el}\right] = \frac{\varepsilon^\mathrm{s,bulk}-\varepsilon[n_{\rm el}]}{\varepsilon^\mathrm{s,bulk}-1} \quad .
\end{equation}
The finite difference in eq.~(\ref{quantum_surface}) is numerically evaluated through a finite difference parameter $\Delta$. In the present work this parameter is set to a low value of $10^{-8}$, with negligible effect of variations around this value on the reported solvation free energies.

The free parameters $\{(\alpha+\gamma),\beta,n_\mathrm{min},n_\mathrm{max}\}$ determining the solvation cavity and the non-mean-field free energy contribution of solvent-solute interactions were optimized by Andreussi \textit{et al.} to give good agreement of solvation energies of a large test set of neutral molecules with experimental values.\cite{Andreussi2012} In this work their parameter values will be used without further optimization, cf. Section~\ref{sec:results-Esolv} below. In addition to these parameters our MPB + Stern layer solvation model contains the ion-specific set of parameters $\{a,d_{\alpha_{\rm ion}},\xi_{\alpha_{\rm ion}}\}$. In total the model depends thus on seven parameters. Defering a systematic optimization of the entire parameter set to a forthcoming publication, we highlight in Section~\ref{sec:activ} how the various ion-specific parameters affect the calculated activity coefficient of KCl aqueous solutions. This in turn also shows that these parameters may be derived by fitting to tabulated salt activity coefficients.

\section{Function-space based MPB solution scheme}
\label{sec:methods}

In the following we develop a function-space based solution scheme of the MPBE, eqs.~(\ref{eq:PBE}) and (\ref{eq:MPBE}), that is adapted to the specificities of a full-potential DFT code like FHI-aims\cite{Blum2009}. FHI-aims expands the Kohn-Sham wavefunctions in highly efficient atom-centered NAO basis functions of the form $\varphi_{{\rm at},i}(\bm r) = \frac{u_{{\rm at},i}(r_\mathrm{at})}{r_\mathrm{at}} Y_{lm}(\Omega_{\rm at})$, where $r_\mathrm{at}=|\bm{r}-\bm{R_\mathrm{at}}|$ and $\Omega_{\rm at}=(\theta_\mathrm{at},\phi_\mathrm{at})$ denote the distance and the angular coordinates of point ${\bm r}$ with respect to the atom ``at'', respectively.  $Y_{lm}(\Omega_{\rm at})$ are the spherical harmonics and $u_{{\rm at},i}(r_\mathrm{at})$ are numerically tabulated and fully flexible radial functions. The resulting real-space integrations are performed on logarithmically-spaced radial grids in order to optimally resolve the Coulomb singularity and orbital cusps close to the atom centers. This irregular grid severely hampers solving the MPBE using common multi-grid finite difference or finite element schemes as used by many PB-DFT schemes\cite{Jinnouchi2008,Fang2013} due to the high costs of interpolation onto regular meshes. But also the present singularities and cusps in an all-electron treatment can contribute to higher computational cost of the common methods. Regular meshes will require very small step sizes and a corresponding large number of nodes, in order to resolve the regions close to the nuclei. Unstructures meshes instead will require an {\em a priori} grid generation step, which, for large problems, can easily become the bottleneck. This is further complicated as the ion exclusion  function and the dielectric function vary rapidly close to cavity's boundary, which itself changes during the SCF-cycle. In contrast, our function-space based solution scheme using multipole representations is specifically designed for the mentioned radial integration grids. We thereby automatically resolve the rapid variation close to the nuclei and also avoid the interpolation between two very different grids. Furthermore, we can exploit the efficient machinery for operating on multipole representations built-in into FHI-aims. Analytical gradients of the free energy functional with respect to the nuclear coordinates, while not within the scope of this study, can in principle be expressed in terms of derivatives of the atom-centered basis functions and the multipole moments of the electrostatic potential.

Referring to the detailed account provided in ref. \cite{Blum2009} FHI-aims achieves an efficient solution of the plain Poisson equation through a regularization of the electrostatic potential $v$ by subtraction of a superposition of free-atom electrostatic potentials $v^{\rm free}_{\rm at}$ derived from non-spinpolarized spherical free-atom electron densities $n^{\rm free}_{\rm el,at}$
\begin{equation}
\delta v = v - \sum_{\rm at} v^{\rm free}_{\rm at} = v - v^\mathrm{free} \quad .
\label{regularization}
\end{equation} 
Both $v^{\rm free}_{\rm at}$ and $n^{\rm free}_{\rm el,at}$, as well as $n^{\rm free}_{\rm el} = \sum_{\rm at} n^{\rm free}_{\rm el,at}$ are accurately known as cubic spline functions on the dense logarithmic grids. The remaining difference potential $\delta v$ is both smooth and free of singularities. It is thus conveniently expanded in multi-center multipole moments\cite{Blum2009}
\begin{equation}
\delta v = \sum\limits_\mathrm{at}\sum\limits_{l=0}^{l_\mathrm{max}}\sum\limits_{m=-l}^l \delta v_{\mathrm{at},lm}(r_\mathrm{at}) Y_{lm}(\Omega_{\rm at}) \quad ,
\label{eq:mpe_basic}
\end{equation}
where $l_\mathrm{max}$ is the maximum angular momentum of the truncated multipole expansion. $\delta v_{\mathrm{at},lm}$ denote the atom-centered multipole components of the difference potential, that are suitably localized on the integration grid around atom ``at'' through a partitioned integration formalism involving partition functions $p_{\rm at}$.\cite{Blum2009} In general all numeric integrations in the FHI-aims package are performed via such a partitioning. The Poisson equation is finally solved in this representation by exploiting the analytic Laplace expansion of the unscreened Green's function $G_0(|\bm{r}-\bm{r'}|) = \frac{1}{4\pi|\bm{r}-\bm{r'}|}$ in terms of spherical harmonics.

\subsection{Newton solver for the MPBE}
\label{sec:Newton}
The formulation of partial differential equations (PDEs) like the MPBE in a function-space oriented solution approach is nowadays quite common.\cite{Holst1995} This is mainly due to the availability of highly efficient solution schemes, like Newton methods, offering fast quadratical convergence.\cite{Deuflhard2004} In our implementation we employ such a Newton solver, albeit not based on commonly used finite element methods\cite{Holst1995}, but rather using a multipole basis expansion which lets us exploit the highly parallel and efficient machinery of FHI-aims without the overhead for mesh generation or uniform grids and any additional interpolation steps.

As a first step we thus reformulate the MPBE, eqs.~(\ref{eq:PBE}) and (\ref{eq:MPBE}), as a functional root-finding problem with respect to $v$
\begin{equation}
F[v] = \nabla \cdot \left[\varepsilon \nabla  v\right] + 4\pi (n_\mathrm{sol} + n_{\rm ion}^{\rm MPB}[v]) = 0 \quad .
\label{eq:NewtonInit}
\end{equation}
Here and in the remainder of this section we thereby drop the explicit $n_{\rm el}$-dependence of $\varepsilon[n_{\rm el}]$ and $\alpha_{\rm ion}[n_{\rm el}]$ in the equations for clarity, recognizing that for the MPBE solver only the $v$-dependence matters. Regularizing $v$ as in eq.~(\ref{regularization}) the root with respect to the difference potential $\delta v$ can then be obtained through an iterative Newton method
\begin{equation}
F'[v_n](\delta v_{n+1}-\delta v_{n}) = -F[v_n] \quad ,
\label{eq:NewtonEq}
\end{equation}
where $F'$ is the Fr\'echet derivative of $F$, the existence of which is proven in the SI. Inserting $F$ and $F'$ yields a LPB-type equation, i.e. a linear PDE in the updated difference potential $\delta v_\mathrm{n+1}$ (for the derivation see the SI)
\begin{equation}
\biggl(\nabla \cdot \left[\varepsilon \nabla\right] - h^2[v_n]\biggr)\delta v_{n+1} \;=\; - 4\pi \varepsilon q[v_n] 
\label{eq:NewtonEq2}
\end{equation}
with 
\begin{equation}
h^2[v_n] = \alpha_{\rm ion} \frac{\alpha_{\rm ion}\phi_0-(\phi_0-1)\cosh(\beta z v_n)}{(1-\phi_0+\phi_0 \alpha_{\rm ion}\cosh(\beta z v_n))^2} \varepsilon^{\rm s,bulk} \kappa^2
\end{equation}
and a modified source term
\begin{align}
- 4& \pi \varepsilon q[v_n] = -4\pi \left(n_\mathrm{el} + n_{\rm ion}^{\rm MPB}[v_n] - \varepsilon n^\mathrm{free}\right) \nonumber\\&- \varepsilon (\nabla \ln(\varepsilon))\cdot(\nabla v^\mathrm{free}) - h^2[v_n]\delta v_n \quad .
\label{eq:newton_1}
\end{align}

Straightforward solution of this LPB-type equation can be achieved by rewriting it in form of a screened Poisson equation (SPE)
\begin{equation}
\biggl(\Delta  - \kappa^2 \biggr)\delta v_{n+1} \;=\; -4\pi q[v_n] + \hat{L}_1[v_n] \delta v_{n+1}
\label{eq:lpbe_mpbe_scf}
\end{equation}
with the response operator
\begin{equation}
\hat{L}_1[v_n] \;=\ -(\nabla \ln(\varepsilon))\cdot\nabla -\left(\kappa^2-\frac{h^2[v_n]}{\varepsilon} \right) \quad .
\label{eq:newton_2}
\end{equation}
The equations \ref{eq:NewtonEq2} or \ref{eq:lpbe_mpbe_scf} could in principle be discretized using one of the standard techniques, such as finite differences or finite elements. The resulting linear algebraic system needs then to be solved numerically, usually employing an iterative solver. A common prerequisite is a suitable preconditioner for the linear system, which will reduce the number of iteration steps. Here, we instead follow a different strategy and perform the preconditioning directly on the function space level.

In principle, a preconditioner can be regarded as an approximation to the inverse of the operator defining our linear problem. In eq.~(\ref{eq:lpbe_mpbe_scf}) the latter is simply $(\Delta  - \kappa^2 - \hat{L}_1[v_n])$, and for $(\Delta  - \kappa^2)$ we know the inverse, which is simply determined by the screened Green's function ${G(|\bm{r}-\bm{r'}|) = \frac{1}{4\pi|\bm{r}-\bm{r'}|}\mathrm{e}^{-\kappa|\bm{r}-\bm{r'}|}}$. Using the Green's function for preconditioning, we multiply eq.~(\ref{eq:lpbe_mpbe_scf}) with $G(|\bm{r}-\bm{r'}|)$ and integrate over space to arrive at
\begin{align}
\delta v_{n+1}(\bm{r}) &= -\int \mathrm{d}\bm{r'} G(|\bm{r}-\bm{r'}|) \biggl(-4\pi q[v_n(\bm{r'})] \nonumber\\&+ \hat{L}_1[v_n(\bm{r'})] \delta v_{n+1}(\bm{r'})\biggr) \quad ,
\label{eq:greens_int1}
\end{align} 
with surface terms vanishing due to the boundary conditions applied on the potential ($v\rightarrow 0$ for $|\bm{r}|\rightarrow\infty$). For the special case $\hat{L}_1[v_n] = 0$ a single evaluation of the right hand side of eq.~(\ref{eq:greens_int1}) would yield $\delta v_{n+1}$ for the next Newton step from the given $v_n$ of the current Newton step. The integration is performed by expanding $\delta v_{n+1}$ in multi-center multipoles as further described in the next subsection. Newton steps are then repeated until convergence in $\delta v$ is reached. 
In contrast, in the general case $\hat{L}_1[v_n] \ne 0$, the right hand side of eq.~(\ref{eq:greens_int1}) also depends on $\delta v_{n+1}$, requiring this equation to be solved self-consistently. 
For that purpose we perform iterative integrations applying our developed multipole expansion relaxation method (MERM). In this method we apply a simple linear mixing scheme with a mixing parameter $\eta$ to the source term $-4\pi q[v_n]+\hat{L}_1[v_n]\delta v_{n+1}$. Thereby, at each Newton step we iteratively solve eq.~(\ref{eq:greens_int1}) for fixed $q[v_n]$ and $\hat{L}_1[v_n]$ until $\delta v_{n+1}$ is converged. This converged $\delta v_{n+1}$ is subsequently used to update $q[v_{n+1}]$ and $\hat{L}_1[v_{n+1}]$ for the next Newton step defining a new SPE to be solved by the relaxation method. As in the special case, Newton steps are then repeated until convergence of $\delta v$ is reached. 

The multi-center multipole expansion for the integrations in eq.~(\ref{eq:greens_int1}) is not a prerequisite, but other approaches can also be employed, in particular any solver for SPEs with fixed right hand sides. Which solver to use can be decided depending on the available infrastructure of the DFT code at hand. Also the iterative linear solver could be replaced by more sophisticated schemes such as Conjugate Gradient\cite{Fisicaro2016}. However, we find that the above approach converges sufficiently fast in all our tests and we do not expect, that the extra amount of CPU load per iteration of higher-level methods will pay off (cf. Section~\ref{sec:performance}).

As a side note, Andreussi \textit{et al.} developed a similar iterative scheme to solve the Poisson equation for a solvent without ions, eq.~(\ref{eq:Poisson}), by their self-consistent continuum solvation (SCCS) scheme\cite{Andreussi2012}. When used with fast Fourier transforms to solve the SPE instead of multipole expansions, our approach formally reduces to the SCCS method for ion-free solvents with $c^{\rm s,bulk} = 0$. In our abstract setting, the SCCS can thereby be considered as making use of the Poisson equation using the unscreened $G_0$ as preconditioner instead of the SPE and the screened $G$. We find the latter favored when solving the PBE or MPBE, as the SPE then describes exactly the bulk limit for ${|\bm{r}|\rightarrow \infty}$, i.e. ${n_\mathrm{ion}^\mathrm{MPB}\rightarrow -\frac{1}{4\pi}\epsilon^{\rm s,bulk}\kappa^2 v}$. This has a beneficial effect on the numerics of the relaxation method as will be addressed in detail in the following section.

\subsection{Solving the screened Poisson equation via multipole expansion}

The main numerical effort of the derived MPB Newton solver lies in solving the SPE by numerical integration of the right hand side of eq.~(\ref{eq:lpbe_mpbe_scf}) via eq.~(\ref{eq:greens_int1}). The developed Newton-MERM scheme is generally applicable and computationally efficient approaches to this integration task will depend on the particular code environment into which the scheme is incorporated. A basic solver for SPEs is in fact often already present in diverse DFT codes\cite{Hasnip2015,Shiihara2008,Blum2009} in the form of Kerker preconditioners\cite{Kerker1981} for the electron density. In case of the Kerker preconditioner utilized by the NAO-based DFT code FHI-aims\cite{Blum2009}, the screened Green's function is first expanded around the atom centers in terms of spherical harmonics similar to the Laplace expansion of the unscreened Green's function \cite{Boschitsch1999,Blum2009}
\begin{align}
G(|\bm{r}-\bm{r'}|) &= \frac{8\kappa}{4\pi} \sum_{\mathrm{at},lm} Y_{lm}(\Omega_\mathrm{at})Y^*_{lm}(\Omega_\mathrm{at}') \nonumber\\&\times
\begin{cases}
 k_l(\kappa r_\mathrm{at})i_l(\kappa r_\mathrm{at}') & \mathrm{for}~r_\mathrm{at}'<r_\mathrm{at}\\
 k_l(\kappa r_\mathrm{at}')i_l(\kappa r_\mathrm{at}) & \mathrm{for}~r_\mathrm{at}<r_\mathrm{at}'
\end{cases} \quad ,
\label{eq:mpe5}
\end{align}
where $i_l$ and $k_l$ are the modified spherical Bessel and Hankel functions, respectively. Inserting this into eq.~(\ref{eq:greens_int1}) yields a multi-center multipole expansion of $\delta v_{n+1}$ analogous to  eq.~(\ref{eq:mpe_basic})
\begin{equation}
\delta v_{n+1} = \sum\limits_\mathrm{at}\sum\limits_{l=0}^{l_\mathrm{max}}\sum\limits_{m=-l}^l \delta v_{\mathrm{at},lm,n+1}(r_\mathrm{at}) Y_{lm}(\Omega_{\rm at}) \quad,
\label{eq:mpe_newton}
\end{equation}
with the corresponding multipole moments $\delta v_{\mathrm{at},lm,n+1}$ given by radial integrals
\begin{align}
\delta& v_{\mathrm{at},lm,n+1}(r_\mathrm{at}) =\nonumber\\ 
&-\frac{8\kappa}{4\pi}\biggl[-4\pi k_l(\kappa r_\mathrm{at})\int\limits_0^{r_\mathrm{at}} \mathrm{d}r_\mathrm{at}' \biggl\{i_l(\kappa r_\mathrm{at}') q_{\mathrm{at},lm}(r_\mathrm{at}')\biggr\} \nonumber\\&+k_l(\kappa r_\mathrm{at})\int\limits_0^{r_\mathrm{at}} \mathrm{d}r_\mathrm{at}'\biggl\{ i_l(\kappa r_\mathrm{at}') \left\{\hat{L}_1\delta v_{n+1}\right\}_{\mathrm{at},lm}(r_\mathrm{at}')\biggr\}\nonumber\\
& -4\pi i_l(\kappa r_\mathrm{at})\int\limits_{r_\mathrm{at}}^{\infty} dr_\mathrm{at}' \biggl\{k_l(\kappa r_\mathrm{at}')  
 q_{\mathrm{at},lm}(r_\mathrm{at}')\biggr\}\nonumber\\&+i_l(\kappa r_\mathrm{at})\int\limits_{r_\mathrm{at}}^{\infty} dr_\mathrm{at}'\biggl\{ k_l(\kappa r_\mathrm{at}')  \left\{\hat{L}_1 \delta v_{n+1}\right\}_{\mathrm{at},lm}(r_\mathrm{at}')\biggr\}\biggr] \quad ,
\label{eq:mpm}
\end{align}
where $q_{\mathrm{at},lm}$ and $\left\{\hat{L}_1 \delta v_{n+1}\right\}_{\mathrm{at},lm}$ are the SPE source term multipole moments obtained by angular integration over the source term itself
\begin{align}
-4&\pi q_{\mathrm{at},lm}(r_\mathrm{at}) \nonumber\\&= -4\pi\int\limits_\mathrm{r_\mathrm{at}} \mathrm{d^2} \Omega_\mathrm{at} \biggl\{p_\mathrm{at}(\bm{r}) q[v_n] Y_{lm}(\Omega_\mathrm{at})\biggr\}
\label{eq:dmp}
\end{align}
and
\begin{align}
&\left\{\hat{L}_1 \delta v_{n+1}\right\}_{\mathrm{at},lm}(r_\mathrm{at}) \nonumber\\&= \int\limits_\mathrm{r_\mathrm{at}} \mathrm{d^2} \Omega_\mathrm{at} \biggl\{p_\mathrm{at}(\bm{r}) \left(\hat{L}_1[v_n] \delta v_{n+1}(\bm r)\right)  Y_{lm}(\Omega_\mathrm{at})\biggr\} \quad .
\label{eq:dmp2}
\end{align}
Numerical evaluation of the radial integral of eq.~(\ref{eq:mpm}) can then efficiently exploit the internal FHI-aims integration grids. Specifically, we employ a multistep Adams-Moulton integrator\cite{Abramowitz1970} and perform an additional interpolation to an extra-fine logarithmic grid with increasing grid shell density close to the nuclei. This allows to optimally resolve the strong variations of $\delta v$ in the vicinity of the nuclei, but involves only a numerically undemanding 1D cubic spline interpolation that is not performance critical compared to the summation of the multipole moments.

In FHI-aims, the extent of the radial basis functions is limited by an atom-specific confinement potential $v_\mathrm{cut,at}$ in order to increase the computational efficiency especially for larger structures.\cite{Blum2009} $v_\mathrm{cut,at}$ smoothly pushes the atomic components of the charge densities $n_\mathrm{el,at}^\mathrm{free}(r_\mathrm{at})$ to zero between the radii $r_\mathrm{onset,at}$ and $r_\mathrm{cut,at}$, confining them spatially to $r_\mathrm{at}<r_\mathrm{cut,at}$. Due to this and the multiplication with the atom-centered partition function $p_{\rm at}$ in eq.~(\ref{eq:dmp}), many contributions to the multipole moments $q_{\mathrm{at},lm}(r_\mathrm{at})$ arising from electron density dependencies in $q[v_n]$ are also automatically confined. 
Confined-source multipole moments are beneficial for computational scaling with system size and speed.\cite{Blum2009} In our approach we therefore also aim to spatially confine all parts of $q_{\mathrm{at},lm}(r_\mathrm{at})$ and $\left\{\hat{L}_1\delta v_{n+1}\right\}_{\mathrm{at},lm}(r_\mathrm{at})$ arising from other terms in eqs.~(\ref{eq:newton_1}) and (\ref{eq:newton_2}). For instance, by choosing a dielectric function of the form of eq.~ (\ref{dielectric_function}) that has a zero gradient outside the transition region many terms in $q_{\mathrm{at},lm}(r_\mathrm{at})$ and $\left\{\hat{L}_1\delta v_{n+1}\right\}_{\mathrm{at},lm}(r_\mathrm{at})$ associated with $\nabla \varepsilon$ terms in eqs.~(\ref{eq:newton_1}) and (\ref{eq:newton_2}) vanish already after the dielectric transition region, which is usually much closer to the nuclei than $r_{\rm onset,at}$. The remaining terms coming from the functions $\kappa^2-\frac{h^2[v_n]}{\varepsilon}$ and $\frac{n_\mathrm{ion}^\mathrm{MPB}[v_n]}{\varepsilon}$ yield multipole moments which become negligibly small already before $r_\mathrm{onset,at}$ for all test cases we looked at, but this detailed convergence with cutoff radius must, of course, be checked for every individual problem, cf. Section~\ref{sec:performance}. The fast decay of the function $\kappa^2-\frac{h^2[v_n]}{\varepsilon}$ is thereby also the main reason why we recast the Newton method into a SPE instead of a Poisson equation as e.g.~done by Andreussi \textit{et al.}\cite{Andreussi2012}. Using a Poisson equation as resolvent would instead give rise to a term $\frac{h^2[v_n]}{\varepsilon}$ in $\hat{L}_1[v_n]$. This term would take a constant value of $\kappa^2$ in the bulk solvent, leading to overall unconfined $\left\{\hat{L}_1\delta v_{n+1}\right\}_{\mathrm{at},lm}(r_\mathrm{at})$ multipole moments. A similar argumentation motivated us not to regularize $v$ with the vacuum potential (cf. eq. (\ref{regularization})) as obtained from a solvent-free calculation in FHI-aims as this also leads to unconfined source multipole moments.

Due to the spatially-confined multipole moments $q_{\mathrm{at},lm}(r_\mathrm{at})$ and $\left\{\hat{L}_1\delta v_{n+1}\right\}_{\mathrm{at},lm}(r_\mathrm{at})$, the explicit radial integration in eq.~(\ref{eq:mpm}) is in principle bounded by the cutoff radius. For grid points $r_{\rm at} < r_{\rm cut,at}$, this means that the integration of the third and fourth term in eq.~(\ref{eq:mpm}) only needs to be carried out up to $r_{\rm cut,at}$. At grid points $r_{\rm at} > r_{\rm cut,at}$ in the far field, the numerical gain is even more pronounced. For such points eq.~(\ref{eq:mpm}) reduces to
\begin{align}
\delta& v_{\mathrm{at},lm,n+1}^\mathrm{ff}(r_\mathrm{at})= \nonumber\\
&-\frac{8\kappa}{4\pi} \Biggl[-4\pi k_l(\kappa r_\mathrm{at})\int\limits_0^{r_\mathrm{at}^\mathrm{cut}} \mathrm{d}r_\mathrm{at}' i_l(\kappa r_\mathrm{at}') q_{\mathrm{at},lm}(r_\mathrm{at}')\nonumber\\&+k_l(\kappa r_\mathrm{at})\int\limits_0^{r_\mathrm{at}^\mathrm{cut}} \mathrm{d}r_\mathrm{at}' i_l(\kappa r_\mathrm{at}') \left(\hat{L}_1\delta v_{n+1}\right)_{\mathrm{at},lm}(r_\mathrm{at}')\Biggr] \quad .
\label{eq:ffmp}
\end{align}
To evaluate $\delta v_{\mathrm{at},lm,n+1}^\mathrm{ff}$, we thus need no additional integration steps in the Adams-Moulton integrator, since the radial integral is independent of $r_{\rm at}$ and therefore fixed for all $r_{\rm at} > r_{\rm cut,at}$. Apart from the obvious numerical gain compared to having to run the Adams-Moulton integrator over a much larger number of grid points, this also implies that the solution of the SPE and therefore also of the MPBE is free of finite integration errors or surface integral terms, since due to the spatially-confined integrand all integrations are formally carried out over the whole space. 
 
Next to these optimized integration routines, the computational efficiency of the iterative multipole-expansion scheme can additionally be improved by exploiting the quick decay of high-$l$ multipole moments in the far field. Similar to the regular multipole-based solution of the Poisson equation \cite{Blum2009}, significant speed-ups and a greatly improved scaling of the MPBE solver can in particular be obtained for large systems by accordingly restricting the actual calculation to low-$l$ multipole moments in the far field. Our implementation of the iterative solver furthermore evaluates the angular and radial integrals associated with $q[v_n]$ only once at the beginning of each Newton step. At each iterative step in the MERM then only integrals associated to $\hat{L}_1[v_n] \delta v_{n+1}$ have to be carried out. Due to the spatial confinement of $\hat{L}_1[v_n] \delta v_{n+1}$ this update is, however, not necessary on the whole integration grid, but instead only on the points where $\hat{L}_1[v_n] \delta v_{n+1} \ne 0$. A full update of $\delta v_{n+1}$ according to eq.~(\ref{eq:mpe_newton}) on the entire integration grid is correspondingly only done after the last iterative MERM step. While generally increasing the computational efficiency, this update strategy is particularly effective for solvent calculations without ions. In this case $\delta v_{n+1}$ has to be updated during the MERM only on the integration grid points of the dielectric transition region and the majority of the integration points close to the nuclei are only considered in the final update.

\subsection{Solution scheme for the LPBE}
\label{sec:lpbe-solver}

The developed multipole-expansion based relaxation method (MERM) is generally suited to solve any LPBEs, since all can in principle be rewritten in form of a SPE. It may therefore also be applied to solve the actual LPBE given in eq.~(\ref{eq:LPBE}). Such a solution could not 
only be of interest for comparison with other PB codes, but also offers a faster alternative to the coupled Newton-MERM solution of the MPBE for cases where the LPBE is a good approximation.

We provide the free energy formula for the LPBE case in the SI. The recasting of the LPBE of eq.~(\ref{eq:LPBE}) into a SPE of the form of eq.~(\ref{eq:lpbe_mpbe_scf}) is analogous to the procedure described in Section \ref{sec:Newton} and leads to a LPB modified source term
\begin{align}
-4&\pi \varepsilon q^\mathrm{LPB} =-4\pi \left(n_{\rm el}- \varepsilon n^\mathrm{free}_{\rm el}\right) \nonumber\\&-\varepsilon\left(\nabla \ln\left(\varepsilon\right)\right)\cdot \left(\nabla v^\mathrm{free}\right)+ \bar{\kappa}^2
v^\mathrm{free}
\label{eq:l1_SPE0}
\end{align}
and LPB response operator
\begin{equation}
\hat{L}_1^\mathrm{LPB} =  - \left(\nabla \ln\left(\varepsilon\right)\right)\cdot\nabla + \left(\frac{\bar{\kappa}^2}{\varepsilon}-\kappa^2\right) \quad .
\label{eq:l1_SPE}
\end{equation}
In terms of numerical accuracy, $q^\mathrm{LPB}$ and $\hat{L}_1^\mathrm{LPB}$ are thereby and in contrast to $q[v_n]$ and $\hat{L}_1[v_n]$ in the Newton scheme, exactly zero beyond the cutoff radius as long as both ionic and dielectric transitions lie inside the confinement region.

\subsection{Multipole correction to the total energy}

It is well known that the introduction of a multipole basis for $\delta v$ leads to a multipole error in the Hartree energy.\cite{Blum2009} In FHI-aims $\delta v$ is regularly obtained by solving the Poisson equation using a multipole-expanded ${\delta n_{\rm el} = n_{\rm el}-n^\mathrm{free}_{\rm el}}$. Expressing the multipole error in $\delta n_{\rm el}$ as ${\delta n^\mathrm{res}_{\rm el} = \delta n_{\rm el} -\delta n^\mathrm{mp}_{\rm el}}$,where $\delta n^\mathrm{mp}_{\rm el}$ is the multipole-expanded regularized electron density and the error in $v$ as ${v^\mathrm{res} = v - v^\mathrm{mp} = \delta v^\mathrm{res}}$, with $v^\mathrm{mp} = v^\mathrm{free} + \delta v^\mathrm{mp}$, where $\delta v^\mathrm{mp}$ is the multipole-expanded regularized electrostatic potential, we can write the interaction energy between electrons and electrostatic potential as
\begin{align}
\frac{1}{2}&\int\mathrm{d}\bm{r}n_{\rm el} v = \int\mathrm{d}\bm{r}n_{\rm el} v^\mathrm{mp} \nonumber\\&- \frac{1}{2}\int\mathrm{d}\bm{r}
\left(n_{\rm el}-\delta n^\mathrm{res}_{\rm el}\right)  v^\mathrm{mp} + \int\mathrm{d}\bm{r}
\delta n^\mathrm{res}_{\rm el} \delta v^\mathrm{res} \quad.
\label{eq:mpcorr_aims}
\end{align}
In FHI-aims, the first term is included in the sum over the eigenvalues in eq.~(\ref{eq:energymin4}) and the second term added as a double counting correction. Neglecting the last term, quadratic convergence in $\delta n^\mathrm{res}_{\rm el}$ can be achieved, as can be seen by expressing $\delta v^\mathrm{res}$ as an integral over $G_0(|\bm{r}-\bm{r'}|)$ and $\delta n^\mathrm{res}_{\rm el}$.\cite{Dunlap1979} 

This multipole error will also exist, when the multipole expansion-based MPBE/LPBE solution scheme is coupled to a DFT code like FHI-aims. The only difference lies in the fact that we do not multipole expand the electron density itself, but the SPE source term. We therefore define the multipole error in the SPE source term, which for the case of the Newton solver is given by
\begin{align}
-4&\pi q^\mathrm{res}+\left\{\hat{L}_1\delta v\right\}^\mathrm{res}\nonumber\\&=-4\pi q + \hat{L}_1\delta v +4\pi q^\mathrm{mp} -\left\{\hat{L}_1\delta v\right\}^\mathrm{mp} \quad ,
\end{align}
where $-4\pi q^\mathrm{mp}$ and $\left\{\hat{L}_1\delta v\right\}^\mathrm{mp}$ are the multipole-expanded SPE source functions. Replacing $\delta n^\mathrm{res}_{\rm el}$ in eq.~(\ref{eq:mpcorr_aims}) with the error in the SPE source term, this allows to write the interaction energy of electron density with electrostatic potential as
\begin{align}
&\frac{1}{2}\int\mathrm{d}\bm{r}n_{\rm el} v =\int\mathrm{d}\bm{r}n_{\rm el} v^\mathrm{mp}\nonumber\\
&-\frac{1}{2}\int\mathrm{d}\bm{r}
\left(n_{\rm el}+\frac{1}{4\pi}\left(-4\pi q^\mathrm{res}+\left\{\hat{L}_1\delta v\right\}^\mathrm{res}\right)\right) v^\mathrm{mp} \nonumber\\&-\frac{1}{4\pi}\int\mathrm{d}\bm{r}
\left(-4\pi q^\mathrm{res}+\left\{\hat{L}_1\delta v\right\}^\mathrm{res}\right) \delta v^\mathrm{res} \quad .
\label{eq:mpcorr_mpbe}
\end{align}
Expressing $\delta v^\mathrm{res}$ in terms of an integral over $G(|\bm{r}-\bm{r'}|)$ and $-4\pi q^\mathrm{res}+\left\{\hat{L}_1\delta v\right\}^\mathrm{res}$, one can again confirm the quadratic convergence of this expression, but this time in the SPE source term multipole error. As required, eq.~(\ref{eq:mpcorr_mpbe}) transfers into eq.~(\ref{eq:mpcorr_aims}) for the solvent-free case with $\varepsilon\equiv 1$ and $\alpha_\mathrm{ion}^\mathrm{MPB}\equiv 0$ everywhere. Equations for the implemented LPB solution scheme, cf. Section \ref{sec:lpbe-solver}, can be derived analogously by replacing the SPE source functions accordingly.

\section{DFT + MPB solver in FHI-aims}

\subsection{Implementation details}

\begin{figure*}
\includegraphics[width=0.8\textwidth]{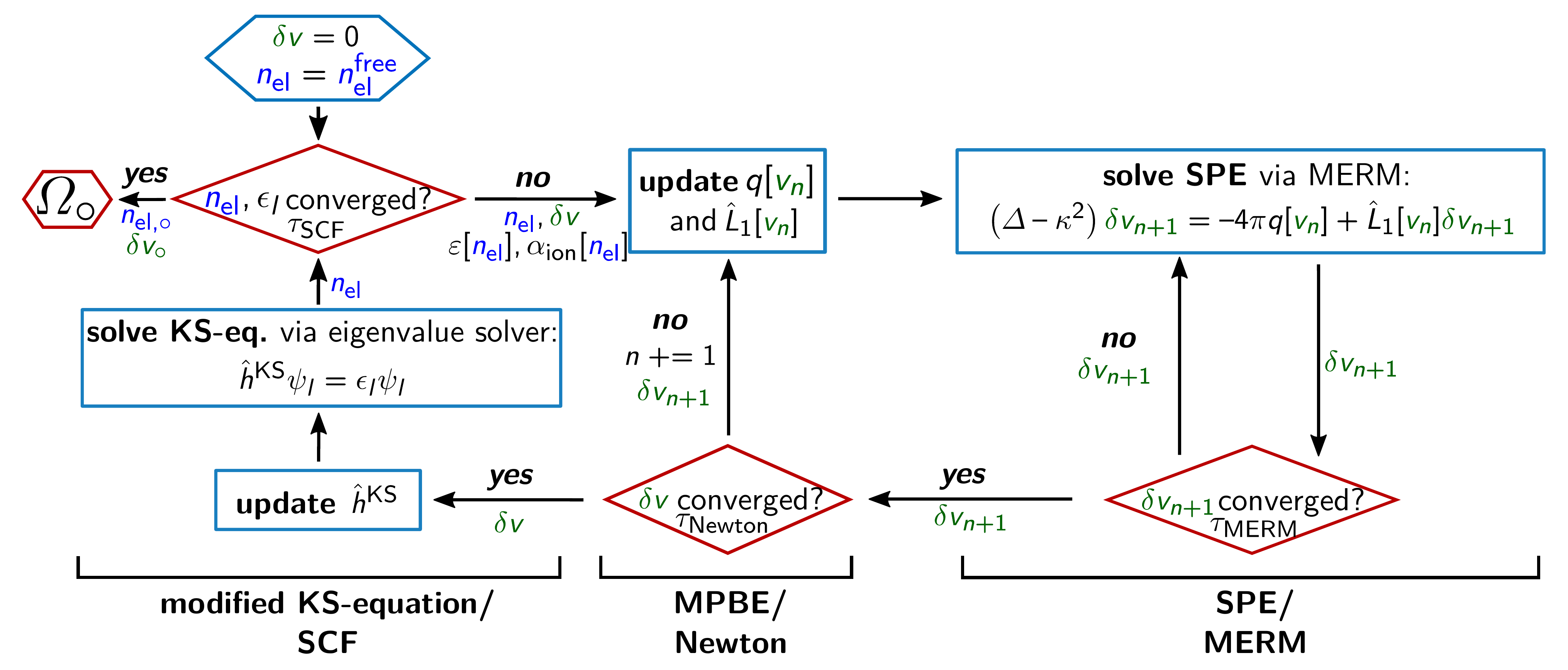}
\caption{Workflow of the Newton-MERM based MPB solver inside FHI-aims. At each electronic SCF step of FHI-aims the iterative Newton scheme to optimize $\delta v$ is initiated. Each Newton step then involves the self-consistent solution of the SPE of eq.~(\ref{eq:lpbe_mpbe_scf}) through the MERM.}
\label{fig:mpbe_scheme}
\end{figure*}

We implemented the Newton-MERM MPB solver into the full-potential DFT code FHI-aims.\cite{Blum2009} Figure \ref{fig:mpbe_scheme} illustrates the basic workflow of this implementation. The workflow for the also implemented MERM-based LPBE solver is analogous and further described in the SI. FHI-aims solves the modified Kohn-Sham equations containing the additional term $\delta v^\mathrm{KS,MPB}_{\varepsilon,\alpha_{\rm ion}}$ of eq.~(\ref{eq:extra_KSHam}) through a SCF cycle. At each corresponding SCF step, i.e. for the then given $n_{\rm el}$, the MPBE of eqs.~(\ref{eq:PBE}) and (\ref{eq:MPBE}) are solved with the Newton-MERM scheme. For this, the iterative Newton scheme to optimize $\delta v$ is initiated, with each Newton step involving the self-consistent solution of the SPE of eq.~(\ref{eq:lpbe_mpbe_scf}) through the MERM. Once the SCF cycle is converged, the resulting ground-state electron density and electrostatic potential are used to evaluate the free energy of the solute $\Omega_0$ in the presence of solvent and ions through eq.~(\ref{eq:energymin4}).

The SCF cycle is initialized with the superposition of free-atom electron densities $n^\mathrm{free}_{\rm el}$ and the superposition of free-atom potentials $v^\mathrm{free}$. In principle, it could be beneficial to start solving the MPBE only after a certain number of SCF steps, i.e. avoid the additional cost of solving the MPBE in the first SCF steps when the electron density still changes largely. However, in practice we obtained a faster SCF convergence in fewer steps when including the MPBE solver directly from the second SCF step onwards. To initiate the MPBE solver, $\varepsilon[n_{\rm el}]$ and $\alpha_{\rm ion}[n_{\rm el}]$ are first evaluated from the $n_{\rm el}$ of the given SCF step. At the very first time the MPBE solver is executed, 

$\delta v$ is initialized with the solution of the corresponding LPBE, cf.~Section~\ref{sec:lpbe-solver}, 
in the case of the MPBE scheme, whereas for LPB calculations, cf.~Section~\ref{sec:lpbe-solver}, we simply set $\delta v = 0$. At all later SCF steps, $\delta v$ is initialized with the self-consistent $\delta v$ of the preceding SCF step. The initialized quantities are used to evaluate the SPE source functions $-4\pi q[v_n]$ and $\hat{L}_1[v_n]\delta v_{n+1}$ (MPBE) or $-4\pi q^\mathrm{LPB}$ and $\hat{L}_1^\mathrm{LPB}\delta v$ (LPBE), and the resulting SPE is solved via the MERM until self-consistency in $\delta v_{n+1}$ (MPBE) or  $\delta v$ (LPBE) is reached. In case of the MPBE solver, the updated $\delta v_{n+1}$ is then used to update the SPE source functions, $n \rightarrow n+1$, and restart the MERM until overall convergence of $\delta v$ is reached. With this converged $\delta v$, the Kohn-Sham Hamiltonian $\hat{h}^\mathrm{KS}$ is updated and the eigenvalue problem is solved in the next SCF step.

FHI-aims measures the numerical convergence of the electron density by evaluating the integrated root mean square change of $n_{\rm el}$ from one SCF step to the next\cite{Blum2009} 
\begin{equation}
\tau_\mathrm{SCF} = \sqrt{\displaystyle\int \mathrm{d}\bm{r} \left\{\left(\delta n_{\rm el,new}({\bf r}) - \delta n_{\rm el,old}({\bf r})\right)^2 \right\}} \quad .
\label{eq:RMSD_n}
\end{equation}
In analogy, the convergence of the Newton method and the MERM is measured through convergence criteria $\tau_\mathrm{Newton}$ and $\tau_\mathrm{MERM}$, respectively, which calculate the corresponding change in the iteratively optimized potential 
\begin{equation}
\tau_\mathrm{Newton/MERM} = \sqrt{\displaystyle\int \mathrm{d}\bm{r} \left\{\left(\delta v_{\rm new}({\bf r}) - \delta v_{\rm old}({\bf r})\right)^2 \right\}} \quad .
\label{eq:RMSD}
\end{equation}

\subsection{Numerical convergence}
\label{sec:performance}

We assess the numerical convergence of the implemented DFT-MPB scheme by calculating a test set of 13 differently functionalized neutral, organic molecules dissolved in water containing a 1M 1:1 electrolyte. This set constitutes a sub-set of the test set introduced by Shivakumar {\em et al.}\cite{Shivakumar2010} and is described in detail in the SI (cf. Table S1). For all calculations we employ the parametrization of Andreussi \textit{et al.} for the dielectric function ($n_\mathrm{min}$ and $n_\mathrm{max}$) and for the non-mean-field solvent-solute interaction part of the solvation energy $\Delta G_{\rm sol}^\mathrm{non-mf}$ ($(\alpha+\gamma)$ and $\beta$) as obtained by their best fit to experimental solvation energies (``fitg03+$\beta$''), see below.\cite{Andreussi2012} For the ion-specific MPBE parameters representative values $a=5$\,{\AA}, $d_{\alpha_\mathrm{ion}}=0.5$, $\xi_{\alpha_\mathrm{ion}}=1$ and $T=300$\,K are used, cf. Section \ref{sec:activ} below. GGA-PBE\cite{Perdew1996} is used as DFT exchange-correlation functional.

\begin{figure}[htb]
\centering
\includegraphics[width=1.\columnwidth]{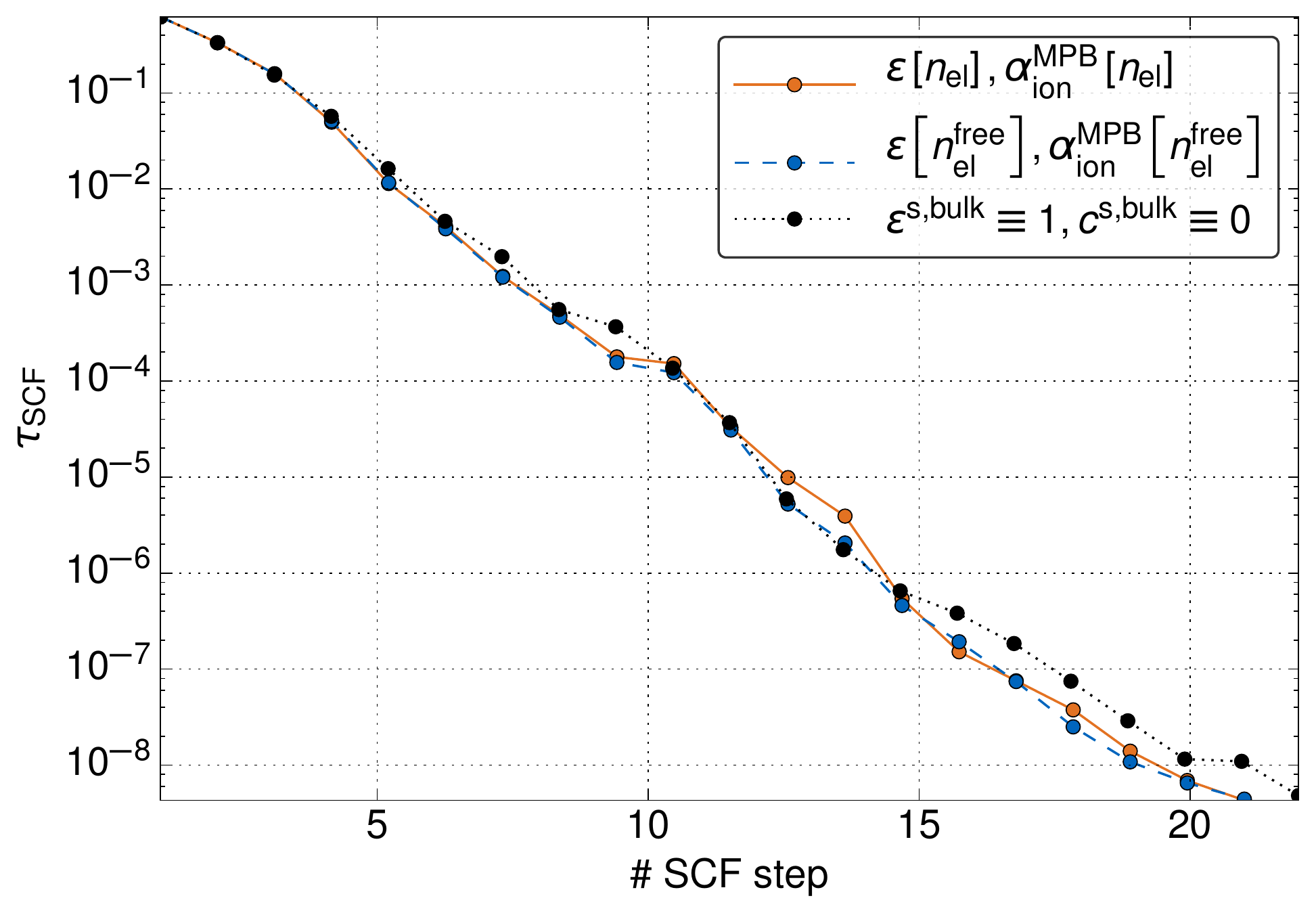}
\caption{Illustration of the SCF convergence for the DFT-MPB scheme in FHI-aims (default ``tight'' settings) using a nitrobenzene molecule dissolved in water containing a 1M 1:1 electrolyte as example, cf. Fig. 1. Compared is the integrated root mean square change of $n_{\rm el}$, $\tau_{\rm SCF}$, as a function of SCF steps for the molecule in vacuum (black dotted line) and in the electrolyte (orange solid line). Additionally shown (blue dashed line) is the convergence when the superposition of free atom densities $n^{\rm free}_{\rm el}$ is used in the evaluation of the dielectric and ion exclusion function for the solvated case, cf. Section \ref{sec:methods-dielec}.}
\label{fig:nitro_scf}
\end{figure}

For this test set, using conservative default convergence criteria of $\tau_\mathrm{Newton/MERM} < 1 \cdot 10^{-10}$ is found to be more than enough to obtain highly converged electrostatic potentials and solvation free energies. Fastest convergence of the MERM is observed for a linear mixing parameter $\eta=0.5$. For this $\eta$ the self-consistent solution of the SPE converged below the convergence criterion for $\tau_\mathrm{MERM}$ is similarly quickly achieved for all tested molecules -- independent of their size and polarity. The number of corresponding MERM steps is initially typically around 60 and then decreases quickly to about five in subsequent Newton and SCF steps, which proves the efficiency of the employed preconditioner. The maximum number of Newton steps required to reach the $\tau_\mathrm{Newton} < 1 \cdot 10^{-10}$ convergence criterion is three. As illustrated in Fig.~\ref{fig:nitro_scf}, at these settings the incorporation of the MPBE solver has only an insignificant effect on the SCF convergence, i.e. the total number of SCF steps required to reach the predefined $\tau_{\rm SCF}$ convergence criterion is about the same with or without the additional solvent calculations. This finding also extends to the case, where the rigid superposition of free atom densities $n^{\rm free}_{\rm el}$ is employed in the evaluation of the dielectric and ion exclusion function, cf. Section \ref{sec:methods-dielec}. 

In general, the computational overhead due to the solvation calculation is therefore mostly defined by the SPE solving step. Therein, the multipole summation is the most expensive computational transformation scaling O$(N^2)$ with system size.\cite{Blum2009} This is, however, drastically improved for larger systems where the confinement of the source multipole moments due to the neglect of all fast-dying far field multipole moments $\delta v_{\mathrm{at},lm,n+1}^\mathrm{ff}(r_\mathrm{at})$ with $l_\mathrm{max}>0$ crucially reduces the computational time of the multipole summation. For smaller systems of the size of those of the molecular test set, this saving doesn't yet kick in, as the integration grid does not extend significantly beyond the cutoff radius. The parallel scalability of our implemented MPBE solver depends thereby entirely on the scaling of the computational bottleneck in form of the multipole summation. Since FHI-aims uses atom-centered integration grids that are designed for optimal parallel scalability, the multipole summations can be very efficiently parallelized, cf. Fig.~(10) in ref. \cite{Blum2009}. More detailed scaling tests fully exploiting parallelization strategies for larger systems will be the topic of a forthcoming publication.

As the MERM scheme performs the numerical integration to solve the SPE on the FHI-aims internal integration grids, it is subject to the same truncation and integration grid parameters already present in any regular FHI-aims DFT calculation. Namely, these are the radial multiplier defining the radial integration grid density (a radial multiplier of $n$ leads to the placing of $n$ additional shells between all present radial shells) , the maximum angular momentum in the multipole expansion $l_{\rm max}$, cf. eqs.~(\ref{eq:mpe_basic}) and (\ref{eq:mpe_newton}), and the radii $r_\mathrm{onset,at}$ and $r_\mathrm{cut,at}$ defined by the confinement potential $v_\mathrm{cut,at}$. The angular grid densities are for very high $l_{\rm max}$ -- if differing from the default settings -- automatically adjusted by FHI-aims to give numerically stable spherical harmonics representations. Convergence of $\Delta G_{\rm sol}$ and $\Delta \Delta G_\mathrm{ion}$ for the molecular test set is obtained at the meV-level for the default ``tight'' production settings for these values, i.e. radial multiplier = 2, $l_{\rm max} = 6$, $r_\mathrm{onset,at} = 4\,${\AA} and $r_\mathrm{cut,at} = 6\,${\AA}. As a consequence, this eliminates the need to introduce separate truncation and integration grid parameters for the MERM. As further detailed in the SI, equivalent findings are obtained for the convergence of $\Delta G_\mathrm{sol}$ and $\Delta \Delta G_\mathrm{ion}$ with the NAO basis, i.e. also here meV-level convergence is obtained at the predefined default settings.

\begin{figure}[htb]
\centering
\includegraphics[width=1.\columnwidth]{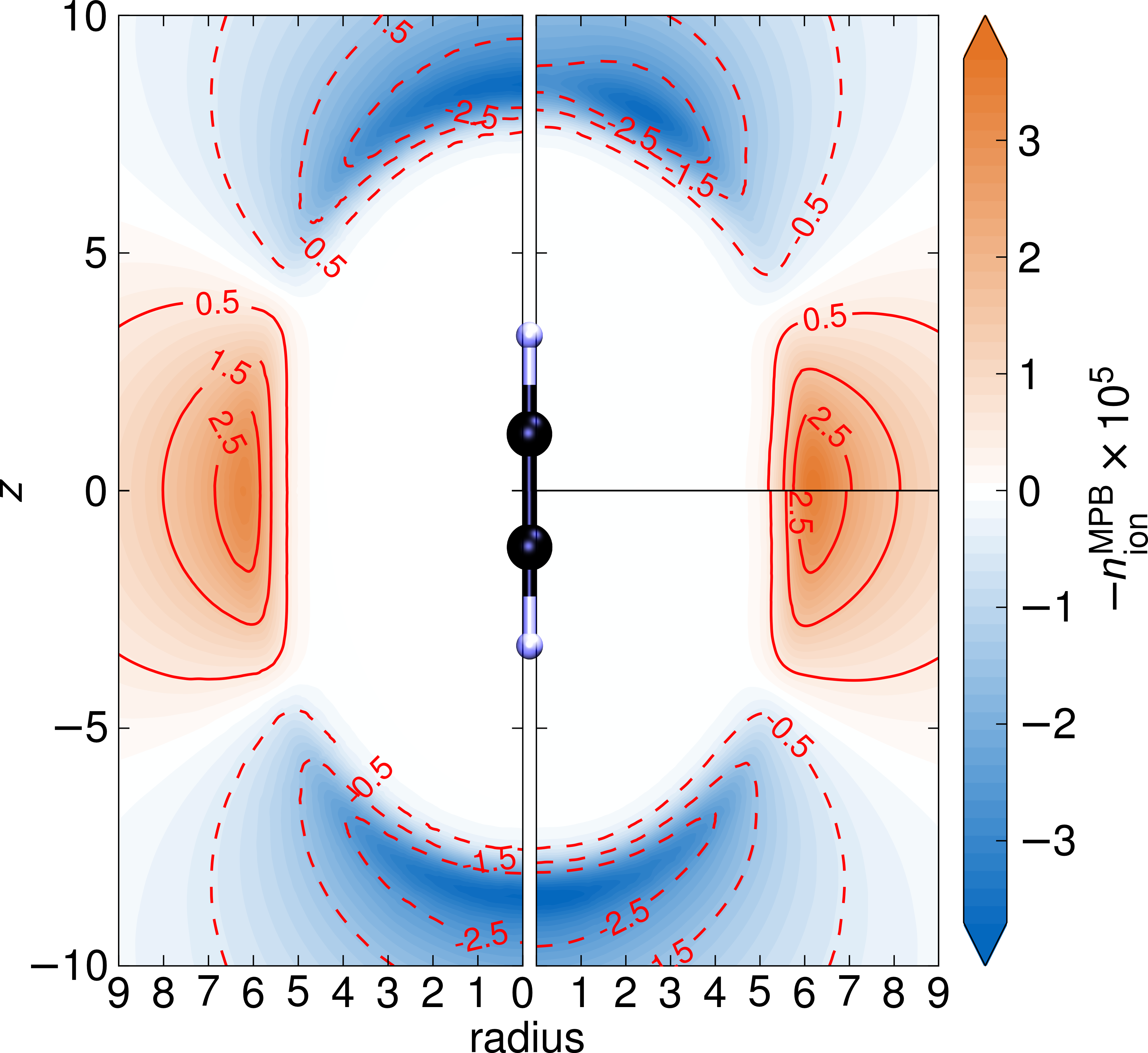}
\caption{Comparison of the ionic charge distribution $n_{\rm ion}^{\rm MPB}$ as calculated for HCCH dissolved in water containing a 1M 1:1 electrolyte (shown with intuitive sign convention), once with the adaptive finite-element code KARDOS (left half) and with the implementation in FHI-aims (right half). The FHI-aims calculations were performed with the default numerical settings $l_\mathrm{max}=6$ (upper panel) and with a higher accuracy of the multipole expansion $l_\mathrm{max}=8$ (lower panel).}
\label{fig:mpbe_lpbe_fem}
\end{figure}

In practice, DFT-MPB solvation free energy calculations can thus be performed at the recommended ``tight'' production settings of FHI-aims.\cite{Blum2009} Further increase of the truncation and integration grid parameters allows to also converge a quantity like the ionic charge distribution $n_{\rm ion}$, which is highly sensitive to small changes in the outer electrostatic potential. To illustrate this we take the ground-state charge distribution for the linear HCCH molecule in vacuum as obtained
at the level of DFT with the GGA-PBE functional\cite{Perdew1996}. For this fixed $n_{\rm el}$, the MPBE is then solved in FHI-aims and in an equivalent external implementation of the MPBE in the adaptive finite element code KARDOS\cite{Lang2013}, again for water containing a 1M 1:1 electrolyte and using the same MPBE parameters as before (for all numerical details of the FEM calculations cf. supplementary information).  Figure \ref{fig:mpbe_lpbe_fem} compares the corresponding results for ionic charge density $n_{\rm ion}^{\rm MPB}$ obtained by by both methods. The left panel shows the results from a highly accurate FEM benchmark calculation, while the upper right panel shows the results from FHI-aims calculation using $l_\mathrm{max}=6$ (default ``tight'' settings) and the lower right panel using $l_\mathrm{max}=8$. As can be seen, qualitative agreement of the ion density is already achieved at FHI-aims production settings. The residual error in the electrostatic potential which we showed to be negligably small on an energy scale (see above) is further reduced by increasing the order of the multipole expansion. Changing to a charged molecule, we even get excellent agreement for the difference of ionic charge densities $n_{\rm ion}^{\rm MPB}-n_{\rm ion}^{\rm LPB}$ which is particularly challenging to resolve and a corresponding plot is shown in the supplementary information.

\begin{figure*}[htb]
	\includegraphics[width=0.8\textwidth]{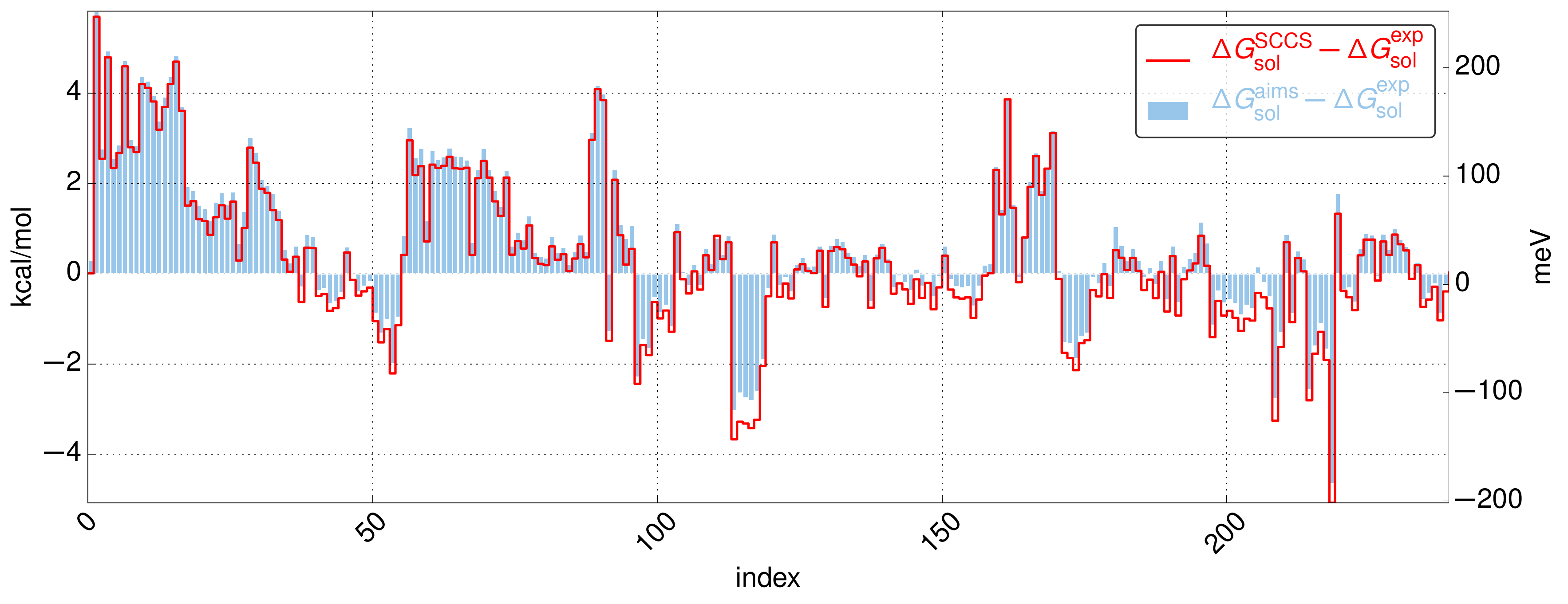}
  \caption{Deviations of calculated room-temperature solvation free energies $\Delta G_\mathrm{sol}(c^\mathrm{s,bulk} = 0)$ from experimental values for the Shivakumar test set of 240 neutral molecules \cite{Shivakumar2010}. Compared are published results from the SCCS solver of Andreussi {\em et al.} \cite{Andreussi2012} with our implementation in FHI-aims (``tight'' settings), both using the optimized ``fitg03+$\beta$'' parameter set. The MAE of the present implementation with respect to experiment is 53\,meV, the MAE with respect to the SCCS solver is 9.3\,meV.}
	\label{fig:aims_vs_marzari}
\end{figure*}

\section{Performance and Parametrization}

\subsection{Solvation free energies of neutral molecules}
\label{sec:results-Esolv}

Like any other implicit solvation method also MPB is an effective approach. Its capabilities and reliability therefore stand and fall with its parametrization. As stated before, our Stern-layer corrected MPB model builds on a total of seven free parameters. These are, on one hand, the two parameters $\{n_\mathrm{min},n_\mathrm{max}\}$ defining the solvation cavity and the two parameters $\{(\alpha+\gamma),\beta\}$ governing the non-mean-field free energy contribution $\Delta G_{\rm sol}^{\rm non-mf}$. On the other hand, there are the three ion-specific parameters $\{a,d_{\alpha_{\rm ion}},\xi_{\alpha_{\rm ion}}\}$ describing the finite ion size and the Stern layer. The latter group of parameters does obviously not enter for ion-free solvents. Aspiring a transferable parameter set that holds for a wide range of systems 
and conditions, this suggests to separately determine and optimize the prior non-ionic parameter group through solvation calculations for ion-free solvents.

This strategy has been pursued by Andreussi \textit{et al.} for their SCCS scheme \cite{Andreussi2012}, to which our MPB + Stern correction scheme reduces formally for ion-free solvents. They optimized the four parameters by fitting to the experimental solvation free energies of the 240-molecule test set of Shivakumar {\em et al.}\cite{Shivakumar2010} to obtain the ``fitg03+$\beta$'' parameter set: ${n_\mathrm{min} = 0.0001}$, ${n_{\rm max} = 0.005}$, $\alpha + \gamma = 50$\,dyn/cm, $\beta = -0.35$\,GPa. Similarly, they arrived at optimized parameter sets for charged solutes, i.e. the ``fit cations'' (${n_\mathrm{min} = 0.0002}$, ${n_{\rm max} = 0.0035}$, $\alpha + \gamma = 5$\,dyn/cm, $\beta = 0.125$\,GPa) and the ``fit anions'' (${n_\mathrm{min} = 0.0024}$, ${n_{\rm max} = 0.0155}$, $\alpha + \gamma = 0$, $\beta = 0.450$\,GPa) parameter sets. Figure~\ref{fig:aims_vs_marzari} reproduces their published solvation free energies $\Delta G_\mathrm{sol}(c^\mathrm{s,bulk} = 0)$ obtained with the ``fitg03$+\beta$'' set for the neutral molecule test set and compares to the corresponding results obtained with our implementation and the same parameter set. Specifically we show the deviation with respect to the experimental reference, and for maximum comparability we employ their reference geometries and the same DFT GGA-PBE functional \cite{Perdew1996} also used by Andreussi {\em et al.}\cite{Andreussi2012}. The agreement between both solvers is excellent with a mean average error (MAE) of 9.3\,meV over the test set. A large part of this small difference is thereby due to the different basis sets employed in the two DFT codes, which affect the position of the solvation cavity via the density cutoffs. Using e.g.~a Gaussian aug-cc-pVDZ basis in the FHI-aims implementation indeed reduces the MAE to an insignificant 6.5\,meV (cf. supplementary information). We emphasize that we validated that this remaining deviation has nothing to do with the non-self-consistent evaluation of the $\Delta G_{\rm sol}^{\rm non-mf}$ contribution in our implementation as compared to the self-consistent evaluation in the SCCS scheme. For the entire test set, this non-mean-field contributions have a negligible effect on the electron density, justifying the computationally efficient treatment of this contribution as a post-correction. In contrast to other authors completely neglecting this contributions\cite{Lespes2015,Steinmann2016}, we expect our approach in general to capture the majority of these effects by at the same time avoiding numerical problems\cite{Steinmann2016}.

\subsection{Activity coefficient of KCl aqueous solutions}
\label{sec:activ}

The true potential of the MPB approach unfolds in the application to electrolyte solutions, where it accounts for effects of a finite ion concentration on top of the pure dielectric response of the solvent. For this, three further ion-specific parameters need to be specified, $\{a,d_{\alpha_{\rm ion}},\xi_{\alpha_{\rm ion}}\}$. Of these three, $a$ describes a finite ion size within the original size-modified PB approach, while $d_{\alpha_{\rm ion}}$ and $\xi_{\alpha_{\rm ion}}$ describe the additional ion exclusion in the Stern layer. All three will generally sensitively determine the description of ionic effects \cite{Chu2007,Harris2014}. Notwithstanding, to date there is no general parametrization protocol that would provide these parameters for a wide range of systems and conditions. Deferring a systematic exploration to a forthcoming publication, we note that activity coefficients of electrolyte solutions could represent an interesting route to this end. Such coefficients are obviously a sensitive measure of ionic effects \cite{Valisko2014,Lenart2007,Slavchov2014,Mester2015} and are tabulated for a wide range of electrolytes \cite{Robinson1959a,Hamer1972}. Fitting to this experimental reference data would then allow to determine the ion-specific MPB parameters, while it would simultaneously provide a first assessment of how well the DFT+MPB+Stern layer approach is capable of treating situations with finite ion concentrations.

We illustrate this concept for KCl aqueous solutions and concentrate on the experimentally well accessible mean molar activity coefficient\cite{Robinson1959}
\begin{equation}
\ln(\gamma^\mathrm{mean}) = \frac{1}{2}\biggl(\ln(\gamma_-) + \ln(\gamma_+)\biggr) \quad ,
\end{equation}
which averages over anionic and cationic contributions. The activity coefficients of anions, $\gamma_-$, and cations, $\gamma_+$, can thereby be expressed as
\begin{align}
k&_\mathrm{B}T\ln(\gamma_\pm(c^\mathrm{s,bulk})) = \mu_\pm(c^\mathrm{s,bulk})-\mu_\pm(c^\mathrm{s,bulk}=0) \nonumber\\&= \Delta \Delta G_\mathrm{ion}^\pm \quad .
\label{eq:activ_freeen}
\end{align}
Here, $\mu_\pm(c^\mathrm{s,bulk})$ and $\mu_\pm(c^\mathrm{s,bulk}=0)$ are the chemical potentials of cation/anion in an electrolyte of salt concentration $c^\mathrm{s,bulk}$ and pure solvent, respectively. Since these chemical potentials represent the free energy change $\frac{\partial \Omega_\circ}{\partial n_\pm}$ of the electrolyte or pure solvent system induced by adding solute charge density $n_\mathrm{sol}$, respectively, the difference of these chemical potentials is just the already introduced ion effect on the solvation free energies $\Delta \Delta G_\mathrm{ion}^\pm$. As a consequence, $\gamma^\mathrm{mean}$ for aqueous solutions containing a certain concentration of monovalent ions such as KCl can thus be obtained from two separate DFT-MPB calculations for an electrolyzed K$^+$ and a Cl$^-$ ion, and the respective two ion-free solvent calculations performed by the DFT-LPB solver. 

\begin{figure}[htb]
\centering
\includegraphics[width=1.\columnwidth]{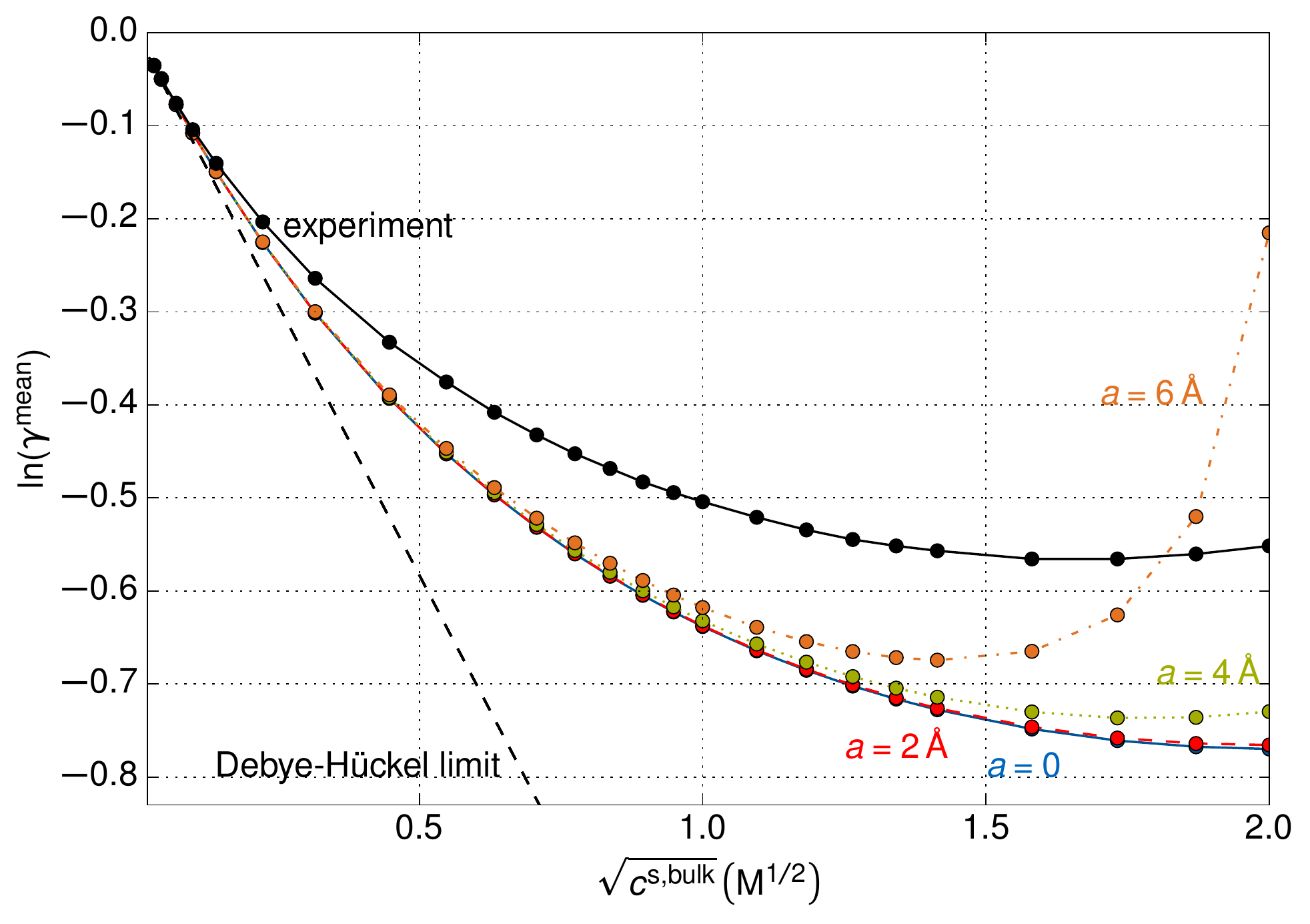}
\caption{Mean molar activity coefficient $\gamma^\mathrm{mean}$ at room temperature as a function of the square root of the ionic bulk concentration $c^{\rm s,bulk}$ of a KCl aqueous solution. The solid black line indicates the experimental curve\cite{Hamer1972}, while the dashed black straight line represents the limit of the Debye-H\"uckel limiting law. Compared are calculated activity coefficients using a range of $a$ values to account for finite size ions. Other parameters are: ``tight'' settings, $\{(\alpha+\gamma),\beta,n_\mathrm{min},n_\mathrm{max}\}$ from the ``fit cations'' and ``fit anions'' parameter sets \cite{Dupont2013}, $d_{\alpha_\mathrm{ion}}=$~0.5, $\xi_{\alpha_\mathrm{ion}}=$~1.0.}
\label{fig:activ_coeff}
\end{figure}

Figure~\ref{fig:activ_coeff} shows corresponding results obtained by using a parameter set of $\{(\alpha+\gamma),\beta,n_\mathrm{min},n_\mathrm{max}\}$ which was optimized for cationic and anionic solutes differing remarkably from the neutral solute parameter set especially for cationic solutes.\cite{Dupont2013} At first, we only tested the option of a finite ion size by choosing parameters $a >0$ and we used a small value for the thickness of the Stern layer, by using $d_{\rm ion} = 0.5$.  Shown is experimental reference data (where we used $m^\mathrm{s,bulk}\approx c^\mathrm{s,bulk}$ for aqueous solutions at room temperature, where $m^\mathrm{s,bulk}$ is the molality of the solution) up to ion concentrations close to the limit of saturated solutions ($m^\mathrm{s,bulk} = 4.803$\,mol/g)\cite{Hamer1972}, as well as the analytic Debye-H\"uckel limiting law
\begin{align}
\ln&(\gamma^\mathrm{mean}(c^\mathrm{s,bulk}))= -\frac{\beta z^2 \kappa}{2 \epsilon^\mathrm{s,bulk}} \nonumber\\&= -1.166 M^{-1/2} \sqrt{c^\mathrm{s,bulk}} \quad ,
\label{eq:DHlimiting}
\end{align}
which is obtained within LPB theory for the purely electrostatic interaction of a point-like charge embedded in a homogeneous dielectric medium with point-like ions of concentration $c_\pm^\mathrm{s,bulk}$. Quite clearly, the deviation of the experimental data from the Debye-H\"uckel limit cannot be accounted for solely on the basis of finite size ions. Only unphysically large values of $a$ yield a significant deviation of the calculated activity coefficient away from the linear Debye-H\"uckel dependence, but do then not produce a curvature that matches the experimental data.

\begin{figure}[htb]
\centering
\includegraphics[width=1.\columnwidth]{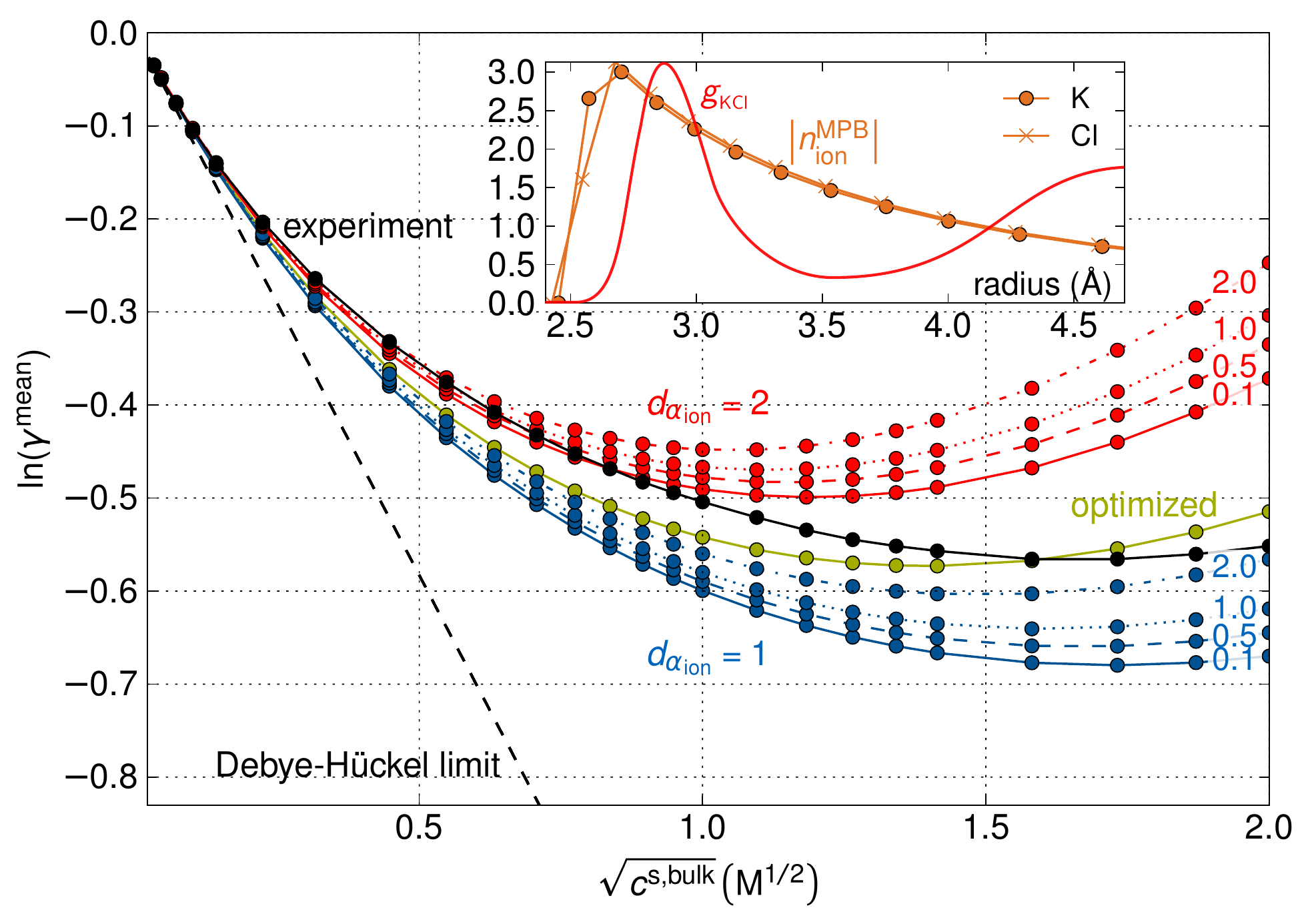}
\caption{Same as Fig.~\ref{fig:activ_coeff}, but this time exploring the effect of different choices of the ionic parameters $d_{\alpha_\mathrm{ion}}$ and $\xi_{\alpha_\mathrm{ion}}$ describing the Stern layer correction. The four upper red curves explore different $\xi_{\alpha_\mathrm{ion}}$ values for $d_{\alpha_\mathrm{ion}} = 2$, the lower blue curves the same for $d_{\alpha_\mathrm{ion}} = 1$. The middle light green curve is the result of an optimum fit to the experimental data achieved for $d_{\alpha_\mathrm{ion}} = 1.54$ and $\xi_{\alpha_\mathrm{ion}} = 0.137$. The inset compares the absolute value of the ionic charge density $n_{\rm ion}^{\rm MPB}$ obtained for the latter parameter values with a typical radial distribution function $g_{\rm KCl}$ obtained from explicit molecular dynamics force field simulations for this system.\cite{Lenart2007} Other parameters as in Fig. \ref{fig:activ_coeff} with $a = 2$\,\AA.}
\label{fig:activ_coeff2}
\end{figure}

This highlights the necessity to consider an additional Stern layer correction in the implicit solvation approach. Figure~\ref{fig:activ_coeff2} correspondingly explores the effect of the thereby introduced ionic parameters $d_{\alpha_\mathrm{ion}}$ and $\xi_{\alpha_\mathrm{ion}}$, which for the present KCl system are identical for the anionic and cationic case. For any significant Stern layer thickness $d_{\alpha_\mathrm{ion}} > 0.5$ we here find the results to be rather insensitive to the exact value of $a$, as had also been reported by Harris {\em et al} \cite{Harris2014}. The results shown in Fig.~\ref{fig:activ_coeff2} are correspondingly obtained for a physically reasonable $a=2$. The calculated activity coefficients vary sensitively with the chosen $(d_{\alpha_\mathrm{ion}}, \xi_{\alpha_\mathrm{ion}})$-pair, indicating that a good account of the experimental variation with ion concentration can be achieved within this two-dimensional parameter space. The light green curve in Fig. \ref{fig:activ_coeff2} demonstrates this for optimized parameter values $d_{\alpha_\mathrm{ion}} = 1.54$ and $\xi_{\alpha_\mathrm{ion}} = 0.137$ as resulting from a simple Nelder-Mead fit to the experimental reference data. For these parameter values the DFT-MPB + Stern layer approach achieves a decent description over a wide range of ionic concentrations, even without any further fine-tuning of the other MPB parameters. For these optimized ionic parameters, the calculated ionic charge density profile $n_{\rm ion}^{\rm MPB}$ around the central ion shows furthermore a good coincidence of the Stern layer onset with the location of the first solvation shell as derived from explicit solvation molecular dynamics simulations by Lenart {\em et al.} \cite{Lenart2007}, cf. inset in Fig.~\ref{fig:activ_coeff2}. This provides some physical legitimation to the parameters and suggests that fitting to experimental activity coefficients provides indeed a reasonable route to a systematic parametrization protocol.

\section{Summary and Conclusions}

We presented a new ansatz to solve the size-modified Poisson-Boltzmann equation for the implicit inclusion of electrolytic solvation effects 
into DFT calculations. The method differs from earlier MPBE solvers in that it employs (screened) Green's functions as preconditioner in a function space setting. Thereby, it can exploit the routines from the DFT code at hand and needs no specialized grids for the solution of the MPBE. For the showcase implementation into the numeric-atomic orbital full-potential DFT code FHI-aims we demonstrated this by combining the Newton solver with a multipole expansion relaxation method that maximally exploits the atom-centered integration grids of this code. For selected test systems excellent agreement with high-accuracy adaptive finite element reference calculations was achieved for ionic charge densities without the need to significantly increase the numerical truncation and integration grid parameters in FHI-aims. Notwithstanding, at present the implicit solvation calculations still impose a noticeable overhead for semi-local DFT calculations of modestly sized molecules, but for larger system sizes and/or higher-rung functionals this relative cost is increasingly reduced, also thanks to the efficient exploitation of the parallelized integration routines offered by FHI-aims.

For ion-free solvents the developed MPB model reduces formally to the self-consistent continuum solvation (SCCS) scheme introduced by Andreussi {\em et al.}\cite{Andreussi2012} We achieve excellent agreement with this scheme for computed solvation free energies for the 240 neutral molecule test set of Shivakumar {\em et al.}\cite{Shivakumar2010} Use of the optimized SCCS parameter sets therefore allows to immediately employ our scheme for transferable solvation calculations for ion-free solvents. However, the true power of the MPB approach unfolds, of course, in the application to electrolytes. We show that after the inclusion of a Stern layer correction the MPB approach is capable of describing the activity coefficient of KCl aqueous solutions over a wide range of ionic concentrations. The sensitivity of the calculated coefficient on the ion parameters employed in the MPB calculation thereby suggests that fitting to extensively tabulated experimental salt activity coefficients could constitute a powerful and general parametrization protocol. We will explore this route in future work to generate an ideally largely transferable parameter set that enables MPB+Stern layer implicit solvation calculations for a wide range of elements and conditions.

\section{Acknowledgments}
The authors gratefully acknowledge support from the Solar Technologies Go Hybrid initiative of the State of Bavaria and the German Science Foundation DFG (grant no. RE1509/21-1). 
S.M.'s research is carried out in the framework of {\sc Matheon} supported by Einstein Foundation Berlin.
We further would like to thank Dr.~Oliviero Andreussi and Prof.~Nicola Marzari for insightful discussions and access to their solvation test set.\\

\textbf{Supporting Information Available}

Derivations of the free energy formulas and Kohn-Sham Hamiltonian correction terms both for the MPBE and the LPBE case are provided. In addition, a detailed description of the applied Newton method is appended along with extensive accuracy and convergence tests of the implemented method.
This material is available free of charge via the Internet at
\url{http://pubs.acs.org/}.

This document is the unedited Author’s version of a Submitted Work that was 
subsequently accepted for publication in the Journal of Chemical Theory and Computation, copyright \copyright American 
Chemical Society after peer review. To access the final edited and published 
work see \url{http://pubs.acs.org/doi/abs/10.1021/acs.jctc.6b00435}.

\bibliography{bibliography}

\end{document}